\documentclass[onecolumn,prd,aps,amsmath,amssymb]{revtex4}
\usepackage{graphicx}
\usepackage{color}
\usepackage{amsmath,amssymb}
\usepackage{array}

\newcommand{\lessim}{\hspace{0.3em}\raisebox{0.4ex}{$<$}\hspace{-0.75em}\raisebox{-.7ex}{$\sim$}\hspace{0.3em}}

\newcommand{\GeV}{\, \text{GeV}}

\newcommand{\tr}{\mbox{tr}}
\newcommand{\gtwo}{I\kern-.1em I\,}
\newcommand{\beq}{\begin{eqnarray}}
\newcommand{\eeq}{\end{eqnarray}}
\newcommand{\bpm}{\begin{pmatrix}}
\newcommand{\epm}{\end{pmatrix}}
\newcommand{\cl}{\, \rm C.L.}

\begin{document}
\title{Constraining dynamical electroweak symmetry breaking via $R_b$} 

\author{Hidenori S. Fukano}
\email{hidenori.f.sakuma@jyu.fi} 
\author{Kimmo Tuominen}
\email{kimmo.i.tuominen@jyu.fi}
\affiliation{Department of Physics, University of Jyv\"askyl\"a, P.O.Box 35, FIN-40014 Jyv\"askyl\"a, Finland \\
and Helsinki Institute of Physics, P.O.Box 64, FIN-00014 University of Helsinki, Finland}

\begin{abstract} New strong gauge interactions remain a viable source for the electroweak symmetry breaking. However, addressing the generation of fermion masses remains a challenge. A basic observable which provides stringent constraints on the flavor extensions of Technicolor-type models is the decay rate of the Z boson into a $b\bar{b}$-pair. In this paper we provide a general framework to evaluate the resulting constraints on the technicolor theory level taking into account the contributions from the vector and axial vector mesons and discuss the consequences for phenomenology.
\end{abstract}

\maketitle

\section{Introduction}

The Standard Model (SM) of the elementary particle interactions, while describing well all current experimental data including the recent discovery of a new boson at the LHC \cite{:2012gk,:2012gu} , is believed to be an incomplete theory. This is mainly so due to its inability to explain the origin of the observed mass patterns of the matter fields, the number of matter generations and why there is excess of matter over antimatter. Several model frameworks beyond SM exist, and one possible paradigm is  to apply strong coupling gauge theory dynamics. 
Technicolor (TC) was originally proposed in \cite{Weinberg:1975gm} (for reviews, see \cite{Hill:2002ap,Sannino:2009za}). In TC the electroweak symmetry breaking is due to the condensation of new matter fields, the technifermions. The old fashioned but simple TC model  based on the QCD-like gauge theory dynamics is incompatible with the electroweak precision data from the LEP experiments \cite{Peskin:1990zt}, and most of the modern model building within the Technicolor paradigm concentrates on the so called walking Technicolor (WTC) \cite{Holdom:1984sk,Yamawaki:1985zg}. Here the Technicolor coupling constant evolves very slowly over a large scale hierarchy due to a nontrivial quasi stable infrared fixed point \cite{Banks:1981nn}. Models of WTC with minimal new particle content can be constructed by considering technifermions to transform under higher representations of the TC gauge group \cite{Sannino:2004qp,Dietrich:2005jn}. 

Technicolor only explains the mass patterns in the gauge sector of the SM via strong dynamics at the electroweak scale $\Lambda_{\textrm{TC}}\simeq {\cal{O}}(1)$ TeV. To explain various mass patterns of the known matter fields within a TC framework, further dynamical mechanism are needed. A well known example is the extended TC (ETC) \cite{Dimopoulos:1979es}, in which the technifemions and the SM fermions are embedded into a larger gauge group (${\cal G}_{\rm ETC}$).  At some high scale, $\Lambda_{\textrm{ETC}}\gg\Lambda_{\textrm{TC}}$, the symmetry ${\cal G}_{\rm ETC}$ is assumed to break down to the TC gauge group. As the technifermion condensation is triggered by the TC gauge dynamics, the SM fermions obtain their masses via the massive ETC gauge bosons coupled with the technifermion condensates. If an ETC gauge group breaks sequentially, such model may explain the observed mass hierarchies of the SM fermions \cite{Appelquist:1993sg,Appelquist:2003hn}. However, it is hard to explain a large top quark mass, or more precisely, a top-bottom mass splitting. To address this particular issue, an alternative to ETC, the top quark condensation model, was proposed in the form of a low energy effective model  \cite{Miransky:1988xi,Nambu89,Marciano89}. Later this model was completed to a topcolor model 
where the gauge group $SU(3)_{\rm QCD} \times U(1)_Y$ of the SM is extended to 
${\cal G}_{\rm topC} = $SU(3)$_1 \times $SU(3)$_2 \times $U(1)$_{Y1} \times $U(1)$_{Y2}$ which is assumed to break at some high scale $\Lambda_{\rm{top}}\gg\Lambda_{\textrm{TC}}$
\cite{Hill:1991at}, and a model combining TC/ETC and topcolor dynamics has been proposed in \cite{Hill:1994hp}, and several groups are pursuing model building along this line \cite{Fukano:2008iv,Ryttov:2010fu,Fukano:2011fp}. 

One of the main experimental constraints on TC/ETC and topcolor arises from the $Z$ boson decay rate to $b\bar{b}$ pairs. More precisely, one considers 
$R_b \equiv \Gamma(Z\to \bar{b}b)/\Gamma(Z\to {\rm had})$.
The importance of various contributions to this observable is determined by the relevant energy scale associated with different stages of the underlying dynamics: The effects from ETC gauge bosons are suppressed by the ETC scale $\Lambda_{\textrm {ETC}}\gg \Lambda_{\textrm{TC}}$, and similarly for the effects of the extended gauge interactions due to the topcolor dynamics. However, the effects from extra goldstone bosons due to topcolor, so called top-pions, are governed by the electroweak scale rather than the topcolor scale. It has been shown that their effect generally is a substantial reduction of $R_b$ relative to the SM prediction and hence this provides stringent constraints on topcolor dynamics \cite{Burdman:1997pf}.

In this paper we consider how these constraints appear in TC models. We will consider the case without any extension towards the matter sectors of SM. The resulting contributions are nontrivial already at this stage, since any TC model features composite vector and axial vector states in the spectrum, and these will mix with the SM gauge fields. Concretely, we will consider a generic low energy effective theory corresponding to the symmetry breaking pattern SU(2)$_L\times$SU(2)$_R\rightarrow$SU(2)$_V$; a concrete walking TC model where this symmetry breaking patter appears in the Next to Minimal Walking TC \cite{Dietrich:2005jn}.

As a low energy effective Lagrangian, we use a Lagrangian based on the generalized hidden local symmety (GHLS) \cite{Bando:1987ym}, in which the dynamical objects are the Nambu-Goldstone bosons (NGBs) and two vector mesons $V,A$ with the EW gauge bosons $W/Z$. 
We are mainly interested in the case where the coupling of the SM fermions and the strongly interacting sector is ``minimal" \cite{Belyaev:2008yj,Andersen:2011yj,Fukano:2009zm,Fukano:2011is} in the sense that the SM fermions couple to the GHLS sector only via the SM gauge fields.
This is in the spirit of the minimal flavor violation \cite{D'Ambrosio:2002ex} in which only the SM interactions break the CP symmetry. The construction of the low energy Lagrangian is detailed in Sec. \ref{lag}.
Then, in Sec. \ref{results}, we compute the $R_b \equiv \Gamma(Z\to \bar{b}b)/\Gamma(Z\to {\rm had})$ assuming minimal coupling with the SM matter fields and study the resulting constraints. 

\section{Lagrangian}
\label{lag}
To construct a low energy effective Lagrangian, we apply the formalism of generalized hidden local symmety (GHLS) \cite{Bando:1987ym,Casalbuoni:1988xm}. The dynamical objects are the Nambu-Goldstone bosons (NGBs) and two vector mesons $V$, $A$ together with the EW gauge bosons $W$, $Z$. We consider the symmetry structure $G_{\rm glo} \times G_{\rm loc}$, where $G_{\rm glo} = SU(2)_L \times SU(2)_R$ and $G_{\rm loc} = SU(2)_L \times SU(2)_R$. The group $G_{\rm glo} \times G_{\rm loc}$ breaks to the diagonal $G_V = SU(2)_V$. 

The choice of unitary gauge would simplify the analysis, since for this symmetry breaking structure all NGBs would be absorbed as longitudinal degrees of freedom of the electroweak gauge bosons, and they would not appear explicitly in the low energy effective theory. However, for the technical requirements of our numerical analysis we need to work in the Feynman gauge, and due to this, we need the interactions between the would-be NGBs and GHLS gauge bosons. In the following, we consider separately the parts containing only GHLS sector fields and the parts containing fermion fields and their interactions.

We decompose the effective Lagrangian as
\beq
{\cal L}_{\rm eff}
=
{\cal L}_0
+
{\cal L}_f 
\label{mFNMWT-Lag}
\,,
\eeq
where ${\cal L}_0$ does not contain quarks and ${\cal L}_f $ is the part including quarks.  We do not consider the leptons since they are inessential for the observables we will analyze in this paper.

\subsection{GHLS part}

Let us begin with the part free of matter fields, ${\cal L}_0$, which contains only GHLS sector fields and is given by 
\beq
{\cal L} 
\!\!\!&=&\!\!\!
{\cal L}_{\rm kin} + a {\cal L}_V + b {\cal L}_A + c {\cal L}_M + d {\cal L}_\pi \,,
\label{GHLS-general}
\eeq
where the kinetic terms are
\beq
{\cal L}_{\rm kin} 
=
-\frac{1}{2} \tr \left[ \tilde{W}^{\mu \nu} \tilde{W}_{\mu \nu}\right] 
-\frac{1}{4} \tilde{B}^{\mu \nu} \tilde{B}_{\mu \nu} 
-\frac{1}{2} \tr \left[ \tilde{V}^{\mu \nu} \tilde{V}_{\mu \nu} + \tilde{A}^{\mu \nu} \tilde{A}_{\mu \nu}\right]\,.
\label{LGHLS-kin}
\eeq
The SU(2) generators are normalized as usual, $\tr(T^aT^b)=\delta^{ab}/2$. The remaining terms are
\beq
&&
{\cal L}_V =
f^2 \tr\left[\left( \hat{\alpha}^\mu_L + \xi_M \hat{\alpha}^\mu_R \xi^\dagger_M \right)^2\right] 
\,, \quad 
{\cal L}_A = 
f^2 \tr\left[\left( \hat{\alpha}^\mu_L - \xi_M \hat{\alpha}^\mu_R \xi^\dagger_M \right)^2 \right] \,,
\label{LGHLS-VA}
\\[1ex]
&&
{\cal L}_M = f^2 \tr\left[\left( \hat{\alpha}^\mu_M \right)^2 \right]
\,, \quad
{\cal L}_\pi = 
f^2 \tr\left[ \left(
\hat{\alpha}_{L \mu} - \xi_M \hat{\alpha}^\mu_R \xi^\dagger_M -\hat{\alpha}^\mu_M \right)^2  \right] \,,
\label{LGHLS-Mpi}
\eeq
where $f$ in Eqs.(\ref{LGHLS-VA}) and (\ref{LGHLS-Mpi}) is a parameter of mass dimension one introduced for dimensional reasons and $a$, $b$, $c$ and $d$ in Eq. (\ref{GHLS-general}) are dimensionless coefficients. 
Each  field $\xi(x)$ transforms under $G_{\rm glo} \times G_{\rm loc}$ as
\beq
&&
\xi_L(x) \to h_L  \xi_L(x)  g^\dagger_L(x)\,, \quad 
\xi_R(x) \to h_R  \xi_R(x)  g^\dagger_R(x) \,, \quad
\xi_M(x) \to h_L(x)  \xi_M(x)  h^\dagger_R(x)\,,
\eeq
where $g_{L(R)}(x) \in [SU(2)_{L(R)}]_{\rm glo}$, $h_{L(R)}(x) \in [SU(2)_{L(R)}]_{\rm loc}$ and $\hat{\alpha}^\mu_{L,R,M}$ denotes the covariant Maurer-Cartan 1-forms which are defined as
\beq
\hat{\alpha}^\mu_{L,R,M} (x)  
\equiv 
\frac{1}{i} \cdot D^\mu \xi_{L,R,M}(x) \cdot \xi^\dagger_{L,R,M}(x)\,.
\eeq
Each covariant derivative is given by
\beq
D_\mu \xi_L(x) 
&=& 
\partial_\mu \xi_L(x) - i L_\mu(x) \xi_L(x) + i \xi_L(x) {\cal L}_\mu(x) \,,
\\[0.5ex] 
D_\mu \xi_R(x) 
&=& \partial_\mu \xi_R(x) - i R_\mu(x)  \xi_R(x) + i \xi_R(x)  {\cal R}_\mu(x) 
\,, \\[0.5ex] 
D_\mu \xi_M(x) 
&=& \partial_\mu \xi_M(x) - i L_\mu(x)  \xi_M(x) + i \xi_M(x)  R_\mu(x) 
\,.
\eeq
Here ${\cal L}_\mu = g \tilde{W}^a_\mu T^a$ and ${\cal R}_\mu = g' \tilde{B}_\mu T^3$  where $\tilde{W},\tilde{B}$ are ordinary electroweak gauge boson fields in  terms of the gauge basis, $g$ and $g'$ are $SU(2)_{\rm EW}$ and $U(1)_Y$ gauge couplings, respectively, and $T^a$ is the $SU(2)$ generator. The fields $L_\mu$ and $R_\mu$ are GHLS gauge bosons associated with  $G_{\rm loc}$, and defined to include the gauge coupling $\tilde{g}$:
\beq
L_\mu = \frac{\tilde{g}\tilde{V}_\mu - \tilde{g}\tilde{A}_\mu}{\sqrt{2}}
\quad , \quad 
R_\mu = \frac{\tilde{g}\tilde{V}_\mu + \tilde{g}\tilde{A}_\mu}{\sqrt{2}}\,.
\eeq

The field strengths in Eq.(\ref{LGHLS-kin}) are then given by
\beq
\tilde{W}_{\mu \nu} &=& 
\partial_\mu \tilde{W}_\nu - \partial_\nu \tilde{W}_\mu 
-ig \left[ \tilde{W}_\mu , \tilde{W}_\nu \right] \,,
\\[1ex]
\tilde{B}_{\mu \nu} &=& 
\partial_\mu \tilde{B}_\nu - \partial_\nu \tilde{B}_\mu \,,
\\[1ex]
\tilde{V}_{\mu \nu} &=& 
\partial_\mu \tilde{V}_\nu - \partial_\nu \tilde{V}_\mu 
-\frac{i \tilde{g}}{\sqrt{2}} \left[ \tilde{V}_\mu , \tilde{V}_\nu \right] 
-\frac{i \tilde{g}}{\sqrt{2}} \left[ \tilde{A}_\mu , \tilde{A}_\nu \right] \,,
\\[1ex]
\tilde{A}_{\mu \nu} &=& 
\partial_\mu \tilde{A}_\nu - \partial_\nu \tilde{A}_\mu 
-\frac{i\tilde{g}}{\sqrt{2}} \left[ \tilde{V}_\mu , \tilde{A}_\nu \right] 
-\frac{i\tilde{g}}{\sqrt{2}} \left[ \tilde{A}_\mu , \tilde{V}_\nu \right] \,.
\eeq

In this paper we parametrize $\xi$ as \cite{Harada:2005br}
\beq
\xi_L = \exp\left[ \frac{i}{\sqrt{2}} \, \phi_L \right] 
\,, \quad
\xi_R = \exp\left[ \frac{i}{\sqrt{2}} \, \phi_R \right] 
\,,\quad
\xi_M = \exp\left[ i \sqrt{2} \phi_M \right]\,,
\label{id-NGBs}
\eeq
where $\phi_{L,R,M}$ are given by
\beq
\phi_L
=
\frac{\tilde{\pi}_\sigma}{f_\sigma} + \frac{\tilde{\pi}_q}{f_q} - (1+\chi) \frac{\tilde{\pi}}{f_\pi}
\,,\quad 
\phi_R
=
\frac{\tilde{\pi}_\sigma}{f_\sigma} - \frac{\tilde{\pi}_q}{f_q} + (1+\chi) \frac{\tilde{\pi}}{f_\pi}
\,,\quad 
\phi_M
=
\frac{\tilde{\pi}_q}{f_q} - \chi \frac{\tilde{\pi}}{f_\pi}\,,
\eeq
and  each decay constant is given by
\beq
f^2_\sigma = 2a\cdot f^2 \,,\quad
f^2_q = 2(b+c) \cdot f^2 \,, \quad 
f^2_\pi = 2 \left( d - c\chi \right) \cdot f^2\,.
\eeq
The parameters $a,b,c,d$ and $f$ are the ones introduced in Eq.(\ref{GHLS-general}), and we define $\chi$ as 
\beq
\chi = \frac{-b}{b+c}\,.
\eeq
To relate with existing literature, we note that $\chi$, as defined above, is the same $\chi$ appearing in \cite{Belyaev:2008yj}, corresponds to $-\zeta$ in \cite{Harada:2005br} and $1-\chi$ in \cite{Foadi:2007ue}. In the present case there are altogether nine would-be NGBs. Among them, three ($\tilde{\pi}^a$) are absorbed by the EW gauge bosons ($W^\pm_\mu/Z_\mu$), three ($\tilde{\pi}^a_\sigma$) are absorbed by the vector mesons ($V^a_\mu$) and three ($\tilde{\pi}^a_q$) are absorbed by the axial-vector mesons ($A^a_\mu$).  Under these paramerizations 
the quantities in Eqs. (\ref{LGHLS-VA}) and (\ref{LGHLS-Mpi}) are represented as  
\beq
\hat{\alpha}^\mu_L + \xi_M \hat{\alpha}^\mu_R \xi^\dagger_M 
\!\!\!&=&\!\!\!
\frac{1}{\sqrt{2}} \partial_\mu (\phi_L + \phi_R) 
- \sqrt{2} \tilde{g} \tilde{V}_\mu + {\cal R}_\mu + {\cal L}_\mu  \nonumber\\
\!\!\!&&\!\!\!
+ \frac{1}{4 i} \left[ \partial_\mu \phi_L , \phi_L\right] 
+ \frac{1}{\sqrt{2} i } \left[ {\cal L}_\mu , \phi_L \right] 
+ \frac{1}{4 i} \left[ \partial_\mu \phi_R , \phi_R\right] 
+ \frac{1}{\sqrt{2} i } \left[ {\cal R}_\mu , \phi_R \right]  \nonumber\\
\!\!\!&&\!\!\!
+
\frac{1}{i}\left[ \partial_\mu \phi_R , \phi_M\right] 
- \frac{\tilde{g}}{i}\left[ \tilde{V}_\mu + \tilde{A}_\mu , \phi_M \right] 
+ \frac{\sqrt{2}}{i} \left[ {\cal R}_\mu, \phi_M \right] + \cdots 
\label{LV-exp}
\,,\\[1ex]
\hat{\alpha}^\mu_L - \xi_M \hat{\alpha}^\mu_R \xi^\dagger_M 
\!\!\!&=&\!\!\!
\frac{1}{\sqrt{2}} \partial_\mu (\phi_L - \phi_R) 
+ \sqrt{2} \tilde{g} \tilde{A}_\mu - {\cal R}_\mu + {\cal L}_\mu  \nonumber\\
\!\!\!&&\!\!\!
+ \frac{1}{4 i} \left[ \partial_\mu \phi_L , \phi_L\right] 
+ \frac{1}{\sqrt{2} i } \left[ {\cal L}_\mu , \phi_L \right] 
- \frac{1}{4 i} \left[ \partial_\mu \phi_R , \phi_R\right] 
- \frac{1}{\sqrt{2} i } \left[ {\cal R}_\mu , \phi_R \right] \nonumber\\
\!\!\!&&\!\!\!
-\frac{1}{i}\left[ \partial_\mu \phi_R , \phi_M\right] 
+\frac{\tilde{g}}{i} \left[ \tilde{V}_\mu + \tilde{A}_\mu , \phi_M \right] 
- \frac{\sqrt{2}}{i} \left[ {\cal R}_\mu, \phi_M \right] + \cdots
\label{LA-exp}
\, \\[1ex]
\hat{\alpha}^\mu_M 
\!\!\!&=&\!\!\!
\sqrt{2} \partial_\mu \phi_M + \sqrt{2} \tilde{g} \tilde{A}_\mu 
+ \frac{1}{i} \left[ \partial_\mu \phi_M , \phi_M\right] 
+\frac{\tilde{g}}{i} \left[ \tilde{V}_\mu + \tilde{A}_\mu , \phi_M \right] + \cdots
\label{LM-exp}
\,, \\[1ex]
\hat{\alpha}^\mu_L - \xi_M \hat{\alpha}^\mu_R \xi^\dagger_M -\hat{\alpha}^\mu_M 
\!\!\!&=&\!\!\!
\frac{1}{\sqrt{2}} \partial_\mu (\phi_L - \phi_R - 2\phi_M) - {\cal R}_\mu + {\cal L}_\mu
\nonumber \\
\!\!\!&&\!\!\!
+ \frac{1}{4 i} \left[ \partial_\mu \phi_L , \phi_L\right] 
+ \frac{1}{\sqrt{2} i } \left[ {\cal L}_\mu , \phi_L \right] 
- \frac{1}{4 i} \left[ \partial_\mu \phi_R , \phi_R\right] 
- \frac{1}{\sqrt{2} i } \left[ {\cal R}_\mu , \phi_R \right] \nonumber\\
\!\!\!&&\!\!\!
- \frac{1}{ i} \left[ \partial_\mu \phi_R , \phi_M\right] 
- \frac{1}{ i} \left[ \partial_\mu \phi_M , \phi_M\right] 
- \frac{\sqrt{2}}{i} \left[ {\cal R}_\mu, \phi_M \right] + \cdots\,.
\label{Lpi-exp}
\eeq
Thus, we decompose ${\cal L}_0$ into 
\beq
{\cal L}_0
= {\cal L}^{(2)}(\{\pi\}, \{ V\}) + {\cal L}^{(3)}(\{\pi\}, \{ V\}) + \cdots\,,
\label{deco-LGHLS}
\eeq
where $\{\pi\}, \{ V\}$ denote collectively $\{\tilde{\pi},\tilde{\pi}_\sigma,\tilde{\pi}_q\}$ and ${\tilde{W}_\mu, \tilde{B}_\mu,\tilde{V}_\mu, \tilde{A}_\mu}$, respectively. 
The terms ${\cal L}^{(i)}$ in Eq.(\ref{deco-LGHLS}) each contain only terms with $i$ fields, i.e. ${\cal L}^{(2)}$ are the quadratic terms, ${\cal L}^{(3)}$ trilinear terms, etc. 
 
In order to see mass terms of NGBs and vector bosons in ${\cal L}^{(2)}$, we substitute Eqs. (\ref{LV-exp}), (\ref{LA-exp}), (\ref{LM-exp}) and (\ref{Lpi-exp}) into ${\cal L}$. Then,  in accordance with \cite{Chivukula:2004mu},  we write ${\cal L}^{(2)}$ as
 \beq
{\cal L}^{(2)} = 
{\cal L}_{\rm kin} + 
\frac{1}{2} \left[ \partial^\mu \vec{\tilde{\Sigma}}^a - ({\cal{Q}} \cdot \vec{\tilde{{\cal G}}}^\mu)^a \right]^T 
 \left[ \partial_\mu \vec{\tilde{\Sigma}}^a - ({\cal Q} \cdot \vec{\tilde{{\cal G}}}_{\mu})^a \right]\,,
 \label{L2-GHLS}
 \eeq
 where $a=1,2,3$ is SU(2) index. The vectors $\vec{\tilde{{\cal G}}}_\mu^a$ and $\vec{\tilde{\Sigma}}^a$ are, in the gauge eigenbasis,
 \beq
 \vec{\tilde{{\cal G}}}^a_\mu
 = \bpm \tilde{B}_\mu \delta^{a3} \,, \tilde{W}^a_\mu \,, \tilde{V}^a_\mu \,, \tilde{A}^a_\mu \epm^T
  \,,\quad
  \vec{\tilde{\Sigma}}^a
 = \bpm \tilde{\pi}^a \,, \tilde{\pi}^a_\sigma \,, \tilde{\pi}^a_q \epm^T\,,
 \label{Gauge-NGB-def}
 \eeq
and ${\cal Q}$ is a $3 \times 4$ matrix given by
\beq
{\cal Q} =
\bpm
\dfrac{-1}{\sqrt{2}}g' f_\pi \delta^{a3} & \dfrac{1}{\sqrt{2}} g f_\pi & 0 & 0 \\[2ex]
\dfrac{-1}{\sqrt{2}}g' f_\sigma \delta^{a3}& \dfrac{-1}{\sqrt{2}} g f_\sigma & \tilde{g} f_\sigma & 0 \\[2ex]
\dfrac{-\chi}{\sqrt{2}}g' f_q \delta^{a3}& \dfrac{\chi}{\sqrt{2}} g f_q & 0& -\tilde{g} f_q 
\epm \,.
\eeq 
As we can see from Eq.(\ref{L2-GHLS}), ${\cal L}^{(2)}$ includes the mixing terms between the NGB fields $\vec{\tilde{\Sigma}}^a$ and the gauge boson fields $\vec{\tilde{{\cal G}}}^a$. In order to eliminate these mixing terms, we add the $R_\xi$ gauge-fixing term:
 \beq
 {\cal L}_{\rm G.F.}
 =
 -\frac{1}{2 \xi} 
 \left[ \partial^\mu \vec{\tilde{{\cal G}}}^a_\mu + \xi \left( {\cal Q}^T \cdot \vec{\tilde{\Sigma}}^a\right)\right]^T \!\cdot
 \left[ \partial^\mu \vec{\tilde{{\cal G}}}^a_\mu + \xi \left( {\cal Q}^T \cdot \vec{\tilde{\Sigma}}^a\right)\right]\,.
\label{LGF-GHLS}
\eeq
As a result, we obtain the mass terms of gauge bosons and NGBs in the gauge basis : 
\beq
{\cal L}^{(2)} + {\cal L}_{\rm G.F.}
=
{\cal L}_{\rm kin}(\{\pi \}, \{ V\})
+
\frac{1}{2} \vec{\tilde{{\cal G}}}_{\mu}^{a T} \!\! \tilde{{\cal M}}^2_{\cal G} \, \vec{\tilde{{\cal G}}}^{a\mu}
-
\frac{1}{2} \vec{\tilde{\Sigma}}^{a T} \!\! \tilde{{\cal M}}^2_\Sigma \, \vec{\tilde{\Sigma}}^a\,,
\label{mass-GHLS}
\eeq
where the mass matrices are represented as
\beq
\tilde{{\cal M}}^2_{\cal G} = {\cal Q}^T\!{\cal Q}\,, \quad
\tilde{{\cal M}}^2_\Sigma = \xi {\cal Q} {\cal Q}^T\,.
\label{VNGB-mass}
\eeq
In this paper, we fix the gauge parameter $\xi=1$ corresponding to the Feynman gauge.
The mass matrices are diagonalized by orthogonal matrices ${\cal O}$ as
\beq
{\cal M}^2_{\cal G} = {\cal O}^T_{\cal G} \tilde{\cal M}^2_{\cal G} {\cal O}^{}_{\cal G}\,, \quad
{\cal M}^2_\Sigma = {\cal O}^T_\Sigma \tilde{\cal M}^2_\Sigma {\cal O}^{}_\Sigma\,.
\eeq
We first define the electric charge eigenstates in the usual manner: e.g. $W^\pm=(W^1_\mu\mp W^2_\mu)/\sqrt{2}$ and similarly for the other states. Then we diagonalize the resulting mass matrices and find 
that  the eigenvalues for the charged bosons are 
\beq
M^2_W = 
\frac{1}{2}g^2f^2_\pi \left[ 1 - \frac{1+\chi^2}{2} \epsilon^2  + {\cal O}(\epsilon^4)\right] 
\,, \quad
 M^2_{V^\pm}  = \tilde{g}^2f^2_\sigma \left[ 1 + \frac{1}{2} \epsilon^2 + {\cal O}(\epsilon^4)\right]
 \,, \quad
 M^2_{A^\pm}  = \tilde{g}^2f^2_q \left[ 1 +  \frac{\chi^2}{2} \epsilon^2 + {\cal O}(\epsilon^4)\right] 
 \,.
\nonumber\\
\label{eigenvalues-GCC}
\eeq
For the neutral bosons we have $M^2_\gamma =  0$ (massless photon) and 
\beq
M^2_Z = 
\frac{g^2 f^2_\pi }{2 c^2_\theta} 
\left[ 1 - \frac{c^2_{2\theta}+\chi^2}{2c^2_\theta} \epsilon^2  + {\cal O}(\epsilon^4)\right]
\,, \quad
 M^2_{V^0} =\tilde{g}^2 f^2_\sigma \left[ 1 + \frac{1}{2c^2_\theta}\epsilon^2 + {\cal O}(\epsilon^4)\right]
 \,,\quad
M^2_{A^0} =
\tilde{g}^2 f^2_q \left[ 1 + \frac{\chi^2}{2c^2_\theta} \epsilon^2 + {\cal O}(\epsilon^4)\right] \,.
\nonumber\\
\label{eigenvalues-GNC}
\eeq
Here we expanded in $\epsilon = g/\tilde{g}$ and $\tan \theta = g'/g$. and denoted $s_\theta \equiv \sin \theta$ and $c_{\theta} \equiv \cos \theta$. Thus, the gauge boson fields and NGB fields in the mass eigenbasis, $\vec{\cal G}^a_\mu\,, \vec{\Sigma^a}$, respectively, are represented as
\beq
\vec{\tilde{{\cal G}}}^a = {\cal O}_{\cal G} \vec{{\cal G}}^a\,, \quad
\vec{\tilde{\Sigma}}^a = {\cal O}_\Sigma \vec{\Sigma}^a \,.
\eeq
The concrete expression of the matrices ${\cal O}_{{\cal G}, \Sigma}$ are given in Appendix \ref{app-diag}.

\subsection{Interactions between SM quarks and GHLS sector }
 
For phenomenology, we need to complete the model by adding the interactions between the SM fermions and would-be NGB fields and vector boson fields. In this paper we  introduce these couplings in the minimal way and do not consider any ETC interactions. This means that $\pi$, which is absorbed by the EW gauge bosons, is the only NGB field that can couple to the SM fermions in terms of the gauge basis; this results in the following Yukawa coupling:
\beq
{ \cal L}_{\Sigma \bar{f}f}
=
- \bar{\psi}_L \left[ 1 + i \frac{\sqrt{2} \tilde{\pi}}{ f_\pi} \right] \!\bpm m_t & 0 \\ 0 & m_b\epm
\psi_R
+ {\rm h.c.}\,,
\label{L-Sff-minimal}
\eeq
where $\psi = (t, b)^T$ is $SU(2)$ doublet and $\psi_{L/R} \to g_{L/R} \psi_{L/R}$ under $G_{\rm glo}$. We note that ${ \cal L}_{\Sigma \bar{f}f}$ breaks $G_{\rm glo}$ symmetry due to the existence of SM-quark mass matrix. In this paper we consider the third family quarks only and we set the $(3,3)$-component of the CKM matrix to unity; $V^{33}_{\rm CKM} = 1$. Moreover, in this minimal way, the SM fermions do not couple with the vector mesons $V,A$ in the gauge basis, i.e. in the gauge basis the gauge interactions of the SM fermions are the usual ones:
\beq
{\cal L}_{G\bar{f}f}
\!\!\!&=&\!\!\! 
gs_\theta \tilde{\gamma}_\mu \bar{\psi} \gamma^\mu \bpm 2/3 & 0 \\ 0 & -1/3 \epm \psi 
+
\frac{g}{\sqrt{2}} \left[ \tilde{W}^+_\mu \bar{t} \gamma^\mu P_L b + {\rm h.c.} \right]
+
\frac{g}{c_\theta} \tilde{Z}_\mu 
\bar{\psi} \gamma^\mu \left[ g_L P_L + g_RP_R\right] \psi
\,,\label{Lff-minimal}
\eeq
where $P_{L,R} \equiv (1\mp \gamma_5)/2$ and $g_{L,R}$ are given by
\beq
&&
g^t_L = \frac{1}{2} - \frac{2}{3} s^2_\theta
\quad , \quad
g^t_R = -\frac{2}{3} s^2_\theta\,,
\label{gZtt}
\\[1ex]
&&
g^b_L = -\frac{1}{2} + \frac{1}{3} s^2_\theta
\quad , \quad
g^b_R = \frac{1}{3} s^2_\theta\,,
\label{gZbb}
\eeq
and $\tilde{\gamma}_\mu, \tilde{W}^\pm_\mu$ and $\tilde{Z}_\mu$ are SM gauge bosons in terms of the gauge basis. 

Thus ${\cal L}$ in Eq.(\ref{mFNMWT-Lag}) is given by
\beq
{\cal L}
=
{\cal L}_{\rm kin} + {\cal L}_{\rm mass} 
+ {\cal L}^{(3)} (\{ \pi\},\{ V\}) 
+ { \cal L}_{\Sigma \bar{f}f} + {\cal L}_{{\cal G}\bar{f}f}\,,
\label{L-mFNMWT-final}
\eeq
where ${\cal L}_{\rm kin}$ and ${\cal L}_{\rm mass}$ are the kinetic term and mass terms for would-be NGBs, vector bosons and quarks. The part ${\cal L}^{(3)}(\{ \pi\},\{ V\})$ contains the trilinear interaction terms of the composite mesons and the parts ${ \cal L}_{\Sigma \bar{f}f}$ and ${\cal L}_{{\cal G}\bar{f}f}$ contain the mass and gauge interactions of the SM fermions. The Lagrangian in Eq.(\ref{L-mFNMWT-final}) contains all operators up to and including dimension four. 

Among the trilinear interaction terms of the vector bosons in Eq. (\ref{L-mFNMWT-final}), we focus on terms which are needed for calculation of the radiative correction to $Z\bar{b}b$-vertex. We denote by ${\cal L}^{(3)}(\{ \pi\},\{ V\})|_{\text{1N}}$ the trilinear interaction terms of the composite mesons involving only one neutral gauge boson, 
\beq
{\cal L}^{(3)}(\{\pi\}, \{ V\})|_{\text{1N}} 
\!\!&=&\!\!
i \sum_{n,a,b} \kappa^n_{ab} 
\left[ 
\tilde{G}^{0 \mu \nu}_{n} \tilde{G}^+_{a\mu} \tilde{G}^-_{b\nu}
+
\tilde{G}^{+ \mu \nu}_a \tilde{G}^-_{b\mu} \tilde{G}^0_{n \nu}
+
\tilde{G}^{- \mu \nu}_lb\tilde{G}^0_{n \mu} \tilde{G}^+_{a \nu}
\right] 
\nonumber\\
&&
+
i g^{\mu \nu} \sum_{n,a,b} g^{n}_{ab} 
\tilde{G}^0_{n \mu}
\left[ \tilde{\pi}^+_a  \tilde{G}^-_{b \nu} -  \tilde{\pi}^-_a \tilde{G}^+_{b \nu}\right]
+
i g^{\mu \nu} \sum_{n,a,b} \lambda^{n}_{ab} \tilde{G}^0_{n\mu}
\left[(\partial_\nu\tilde{\pi}^+_a)\tilde{\pi}^-_b   - (\partial_\nu\tilde{\pi}^-_a) \tilde{\pi}^+_b \right]
\nonumber \\
\!\!&=&\!\!
i \sum_{i,j} \kappa^{(Z)}_{ij} 
\left[ 
Z^{\mu \nu} G^+_{i \mu} G^-_{j \nu}
+
G^{+ \mu \nu}_{i} G^-_{j \mu} Z_\nu
+
G^{- \mu \nu}_{j} Z_\mu G^+_{i \nu}
\right]
\nonumber\\[1ex]
&&
+
i g^{\mu \nu} \sum_{i,j} g^{(Z)}_{ij}  Z^0_\mu \left[  \pi^+_i  G^-_{j \nu} -\pi^-_i  G^+_{j \nu} \right]
+
i g^{\mu \nu} \sum_{i,j} \lambda^{(Z)}_{ij} Z^0_\mu \left[(\partial_\nu\pi^+_i) \pi^-_j  - ( \partial_\nu\pi^-_i)\pi^+_j \right]
+(\cdots)
\,,\label{GHLS-tripleVS}
\eeq
where $\tilde{G}^{\mu \nu}_n $ is defined as $\tilde{G}^{\mu \nu}_n \equiv \partial^\mu \tilde{G}^\nu_n -\partial^\nu \tilde{G}^\mu_n$ and  the first line corresponds to the triple gauge boson interactions which are derived from the kinetic term of vector bosons in Eq. (\ref{LGHLS-kin}), and the second line corresponds to the trilinear interactions which are derived from the GHLS part Eq. (\ref{deco-LGHLS}). The first two lines in Eqs. (\ref{GHLS-tripleVS}) are in terms of gauge  basis, on the other hand, the last two lines are in terms of mass basis. In the last line in Eq.(\ref{GHLS-tripleVS}) the dots $(\cdots)$ show the interactions among NGBs and vector boson fields other than $Z$.  Here $n = 0,1,2,3$ and $a,b,i,j= 1,2,3$ correspond to $\gamma,Z/W^\pm, V^{0,\pm},A^{0,\pm}$ and $\pi,\pi_{\sigma},\pi_q$, respectively. Now, $\kappa^n_{ab}$ are given by
\beq
\kappa^0_{11} = g s_\theta 
\quad , \quad
\kappa^1_{11} = g c_\theta 
\quad , \quad
\kappa^2_{22} = \kappa^2_{33} = \kappa^3_{23} = \kappa^3_{32} = \frac{\tilde{g}}{\sqrt{2}}
\quad , \quad 
\text{others $=0$}\,,
\eeq
and $g^{n}_{\pi W}, \lambda^{n}_{\pi \pi}$ are given in Tables \ref{coupling-VVphi} and \ref{coupling-Vphiphi}. 
Moreover, $\kappa^{(Z)}_{ij}, g^{(Z)}_{ij}$ and $\lambda^{(Z)}_{ij}$ are defined as
\beq
\kappa^{(Z)}_{ij} = \sum_{n,a,b} \kappa^n_{ab}[v^n_Z] [v^a_{G^\pm_i}] [v^b_{G^\pm_j}]
\quad ,\quad
g^{(Z)}_{ij} = \sum_{n,a,b} g^{n}_{ab} [v^n_Z] [v^a_{\pi^\pm_i}] [v^b_{G^\pm_j}]
\quad , \quad
\lambda^{(Z)}_{ij} = \sum_{n,a,b} \lambda^{n}_{ab} [v^n_Z] [v^a_{\pi^\pm_i}] [v^b_{\pi^\pm_j}]
\,,\label{coupling-for-massbasis}
\eeq
where each $[v^a_V]$ is a component in the matrix ${\cal O}$ translating between the gauge- and mass eigenbases. These matrices are given explicitly in Appendix \ref{app-diag}. In the notation $[v^a_A]$, the index $``a"$ corresponds to the vector boson fields in the gauge basis, while the index $``A"$ refers to the vector boson fields in the mass basis.

\begin{table}[htbp]
%
%
{
\tabcolsep=1ex
\renewcommand\arraystretch{1.5}
\begin{tabular}{|c|c|c|c|}
\hline 
$g^0_{ab}$ & $b=1$ & $b=2$ & $b=3$ 
\\ \hline \hline 
$a=1$ & $g^2s_\theta(-f^2_\sigma + f^2_\pi + \chi^2 f^2_q)/(\sqrt{2} f_\pi)$ & $0$ & $\chi g\tilde{g} s_\theta (f^2_\sigma - f^2_q)/f_\pi$ 
\\\hline 
$a=2$ & $0$ & $g\tilde{g}f_\sigma s_\theta$ & $0$ 
\\\hline
$a=3$ & $0$ & $0$ &$-gs_\theta\tilde{g}f^2_\sigma/f_q$ 
\\\hline 
\end{tabular}
}\\[2ex]
%
%
{
\tabcolsep=1ex
\renewcommand\arraystretch{1.5}
\begin{tabular}{|c|c|c|c|}
\hline 
$g^1_{ab}$ & $b=1$ & $b=2$ & $b=3$
\\ \hline \hline 
$a=1$ & $g^2s^2_\theta(f^2_\sigma - f^2_\pi - \chi^2 f^2_q)/(\sqrt{2}c_\theta f_\pi)$ &   $-g\tilde{g}(f^2_\sigma - \chi^2 f^2_q)/(2c_\theta f_\pi)$ & $\chi g\tilde{g}c_{2\theta}(f^2_\sigma - f^2_q)/(2c_\theta f_\pi)$
\\\hline 
$a=2$ & $0$ & $g\tilde{g} c_{2\theta} f_\sigma /(2c_\theta)$ &  $\chi g \tilde{g}f^2_q/(2c_\theta f_\sigma)$
\\\hline
$a=3$ & $0$ & $-\chi g\tilde{g} f_q/(2c_\theta)$& $-g\tilde{g}c_{2\theta}f^2_\sigma/(2c_\theta f_q)$
\\\hline 
\end{tabular}
}\\[2ex]
%
%
{
\tabcolsep=1ex
\renewcommand\arraystretch{1.5}
\begin{tabular}{|c|c|c|c|}
\hline 
$g^2_{ab}$ &  $b=1 $ & $b=2$ & $b=3$
\\ \hline \hline 
$a=1$ & $g\tilde{g}(f^2_\sigma - \chi^2f^2_q)/(2f_\pi)$ & $0$ & $-\chi \tilde{g}^2(f^2_\sigma - f^2_q)/(\sqrt{2} f_\pi)$
\\\hline 
$a=2$ &  $-g\tilde{g}f_\sigma/2$  & $0$ & $0$
\\\hline
$a=3$ & $\chi g\tilde{g} f_q/2$  & $0$ & $\tilde{g}^2(f^2_\sigma - f^2_q)/(\sqrt{2} f_q)$
\\\hline 
\end{tabular}
}\\[2ex]
%
%
{
\tabcolsep=1ex
\renewcommand\arraystretch{1.5}
\begin{tabular}{|c|c|c|c|}
\hline 
$g^3_{ab}$ & $b=1$ & $b=2$ & $b=3$
\\ \hline \hline 
$a=1$ & $-\chi g\tilde{g}(f^2_\sigma - f^2_q)/(2f_\pi)$ & $\chi \tilde{g}^2(f^2_\sigma - f^2_q)/(\sqrt{2} f_\pi)$ & $0$
\\\hline 
$a=2$ & $-\chi g \tilde{g}f^2_q/(2 f_\sigma)$ & $0$ & $0$
\\\hline
$a=3$ & $g\tilde{g}f^2_\sigma/(2 f_q)$ & $-\tilde{g}^2(f^2_\sigma - f^2_q)/(\sqrt{2} f_q)$  & $0$
\\\hline 
\end{tabular}
}
\caption{
$g^{n}_{ab}$ in Eqs. (\ref{GHLS-tripleVS}) and (\ref{coupling-for-massbasis}).
\label{coupling-VVphi}
}
\end{table}

\begin{table}[htbp]
{
\tabcolsep=1ex
\renewcommand\arraystretch{1.5}
\begin{tabular}{|c|c|c|c|}
\hline 
$\lambda^{0}_{ab}$ & $b=1 $ & $b = 2 $ & $b = 3 $ 
\\ \hline \hline 
$a = 1 $ & $gs_\theta [2f^2_\pi - (1-\chi^2)f^2_\sigma]/(2f^2_\pi)$  & $0$ &  $-gs_\theta(1+\chi)f^2_\sigma/(2f_\pi f_q)$
\\\hline 
$a = 2 $ & $0$ & $gs_\theta/2$ & $0$  
\\\hline 
$a= 3 $ & $gs_\theta [(1-\chi)f^2_\sigma + 2 \chi f^2_q]/(2f_\pi f_q)$  & $0$ & $gs_\theta f^2_\sigma/(2f^2_q)$ 
\\\hline
\end{tabular}
}\\[2ex]
%
%
{
\tabcolsep=1ex
\renewcommand\arraystretch{1.5}
\begin{tabular}{|c|c|c|c|}
\hline 
$\lambda^{1}_{ab}$ & $b=1$ & $b = 2$ & $b = 3$ 
\\ \hline \hline 
$a = 1$ & $gc_{2\theta}[2f^2_\pi - (1-\chi^2)f^2_\sigma]/(4c_\theta f^2_\pi)$ & $g[\chi^2(1+\chi)f^2_q - (1-\chi)f^2_\pi]/(4c_\theta f_\pi f_\sigma)$&  $-gc_{2\theta}(1+\chi)f^2_\sigma/(4c_\theta f_q f_\pi)$ 
\\\hline 
$a = 2$ & $-g[2 f^2_\sigma - \chi^2(1-\chi)f^2_q - (1-\chi) f^2_\pi]/(4c_\theta f_\pi f_\sigma)$ & $gc_{2\theta}/(4c_\theta)$ & $g(f^2_\pi + \chi^2 f^2_q)/(4c_\theta f_q f_\sigma)$
\\\hline 
$a= 3$ &  $gc_{2\theta} [(1-\chi)f^2_\sigma + 2 \chi f^2_q]/(4c_\theta f_\pi f_q)$&$-g[f^2_\pi + \chi (\chi + 2) f^2_q]/(4c_\theta f_q f_\sigma)$& $gc_{2\theta}f^2_\sigma/(4c_\theta f^2_q)$  
\\\hline
\end{tabular}
}\\[2ex]
%
%
{
\tabcolsep=1ex
\renewcommand\arraystretch{1.5}
\begin{tabular}{|c|c|c|c|}
\hline 
$\lambda^{2}_{ab}$ & $b=1$ & $b =2$ & $b = 3$ 
\\ \hline \hline
$a = 1$ & $\tilde{g}(1-\chi^2) f^2_\sigma/(2\sqrt{2}f^2_\pi)$ & $0$ & $\tilde{g} (1+\chi)f^2_\sigma/(2\sqrt{2}f_\pi f_q)$
\\\hline  
$a = 2$ & $0$ & $\tilde{g}/(2\sqrt{2})$ & $0$  
\\\hline 
$a = 3$ &  $-\tilde{g}[2\chi f^2_q + (1-\chi)f^2_\sigma]/(2\sqrt{2}f_\pi f_q)$ & $0$ & $\tilde{g}(2f^2_q - f^2_\sigma)/(2\sqrt{2}f^2_q)$  
\\\hline
\end{tabular}
}\\[2ex]
%
%
{
\tabcolsep=1ex
\renewcommand\arraystretch{1.5}
\begin{tabular}{|c|c|c|c|}
\hline 
$\lambda^{3}_{ab}$ & $a=1$ & $a =2$ & $a = 3$ 
\\ \hline \hline
$b = 1$ & $0$ & $-\tilde{g}\chi(1+\chi)f^2_q/(2\sqrt{2}f_\pi f_\sigma)$ & $0$
\\\hline  
$b = 2$ & $\tilde{g}\chi[2 f^2_\sigma -(1- \chi) f^2_q]/(2\sqrt{2}f_\pi f_\sigma)$ & $0$ & $-\tilde{g}(2f^2_\sigma + \chi f^2_q)/(2\sqrt{2}f_\sigma f_q)$
\\\hline 
$b = 3$ & $0$ & $\chi \tilde{g} f_q/(2\sqrt{2}f_\sigma)$ & $0$  
\\\hline
\end{tabular}
}
\caption{
$\lambda^{n}_{ab}$ in Eqs. (\ref{GHLS-tripleVS}) and (\ref{coupling-for-massbasis}).
\label{coupling-Vphiphi}
}
\end{table}

\section{Calculation of $R_b$ and numerical results}
\label{results}

The $Zb\bar{b}$ vertex in the mass basis is
\beq
{\cal L}_{Zbb}
=
Z_\mu \bar{b} \gamma^\mu 
\left( \Gamma^b_L \frac{1-\gamma_5}{2} + \Gamma^b_R \frac{1+ \gamma_5}{2} \right) b
\,.
\eeq
We calculate the corrections to $\Gamma^b_{L,R}$ due to GHLS fields, and we denote these corrections as $\delta \Gamma_{L,R}$. As already emphasized, in this paper we work in the 't Hooft-Feynman gauge and use dimensional regularization. In the calculation of the Feynman diagrams required for the form factors of $Zb\bar{b}$ vertex we also adopt the condition $m^2_b = 0$. 

First, we consider the physical electroweak parameters. Among several ways to choose the independent parameters, we choose the electroweak parameters $v_{\rm EW}\,, g_{\rm EW}\,, g'_{\rm EW}$ by the two point current correlation functions in accordance with \cite{Barbieri:2004qk}.
The electroweak scale is $v_{\rm EW} = (246\GeV)\equiv \sqrt{2}f_\pi$, and the electroweak gauge couplings are given in terms of the gauge couplings and GHLS model parameters as
\beq
\frac{1}{g^2_{\rm EW}} 
\equiv
\frac{1}{g^2}  + \frac{1 + \chi^2}{2\tilde{g}^2} 
\,, \quad
\frac{1}{g'^2_{\rm EW}} 
\equiv
\frac{1}{g'^2}  + \frac{1 + \chi^2}{2\tilde{g}^2} 
\,.
\label{def-EW-phys}
\eeq
The physical Weinberg angle must also be defined carefully. First, we define the QED gauge coupling $e$. Since the $U(1)_{\rm e.m}$ symmetry is conserved, the QED coupling $e$ should be defined as
\beq
e \equiv g s_\theta [v^0_\gamma] 
\,,
\eeq
which arises from the first term in Eq.(\ref{Lff-minimal}) in the mass basis. Note that 
the above definition of the electroweak couplings using the two point current correlation functions does not reproduce exactly the SM-like relation $1/e^2 = 1/g^2_{\rm EW} + 1/g'^2_{\rm EW}$. Hence we define $\sin$ and $\cos$ of the Weinberg angle as a ratio of $e$ to $g_{\rm EW}$ and $g'_{\rm EW}$ , i.e.
\beq
&&
\sin \theta_W \equiv \frac{e}{g_{\rm EW}}
= s_\theta \left[ 1 + \frac{1+\chi^2 - 4s^2_\theta}{4} \epsilon^2 + {\cal O}(\epsilon^4)\right]\,,
\\
&&
\cos \theta_W \equiv \frac{e}{g'_{\rm EW}}
= c_\theta \left[ 1 + \frac{1+\chi^2 - 4c^2_\theta}{4} t^2_\theta \epsilon^2 + {\cal O}(\epsilon^4)\right]\,.
\label{EW-angle}
\eeq
These definitions imply
\beq
\sin^2 \theta_W + \cos^2 \theta_W
=
1 -(1-\chi^2) s^2_\theta \epsilon^2 + {\cal O}(\epsilon^4)\,,
\eeq
which becomes $1$ in the limit $\tilde{g} \to \infty$ or $\chi^2 \to 1$ corresponding to the SM.

Based on the above definitions, after changing from the gauge basis to the mass basis, the tree level couplings $\Gamma^b_{L,R}$ in Eq.(\ref{Lff-minimal})
are given by 
\beq
\Gamma^b_{L,R} \left[ \text{GHLS,tree} \right]
=
- \frac{1}{3} g s_\theta [v^0_Z]
+\frac{g}{c_\theta}g^b_{L,R} [v^1_Z] \,.
\,,\label{gZbb-tree}
\eeq
We express $g$ in terms of the coupling $g_{\rm EW}$ using the relation 
\beq
\frac{g_{\rm EW}}{\cos\theta_W}
=
\frac{g}{c_\theta} \left[ 1- \frac{c^2_{2\theta} + \chi^2}{4c^2_\theta} \epsilon^2 + {\cal O}(\epsilon^4)\right]
=
\frac{g}{c_\theta}[v^1_Z]
\,,\label{gbLL}
\eeq
obtained from (\ref{def-EW-phys}) and valid to order ${\cal O}(\epsilon^2_0)$. 
Also, we insert $g^b_{L,R}$ as given in Eq.(\ref{gZbb}), and obtain
\beq
\Gamma^b_L[\text{GHLS,tree}]
\!\!&=&\!\!
\frac{g_{\rm EW}}{\cos \theta_W} \left[ -\frac{1}{2} + \frac{1}{3} \cdot 
s^2_\theta \left( 1- \frac{c_\theta [v^0_Z]}{s_\theta [v^1_Z]}\right)
\right]
\,,\label{GHLS-treeZbbL}\\[1ex]
\Gamma^b_R[\text{GHLS,tree}]
\!\!&=&\!\!
\frac{g_{\rm EW}}{\cos \theta_W} \cdot  \frac{1}{3} \cdot 
s^2_\theta \left( 1- \frac{c_\theta [v^0_Z]}{s_\theta [v^1_Z]}\right)
\,,\label{GHLS-treeZbbR}
\eeq
where the components of $[v]$ are given in Appendix.\ref{app-diag}.

The one-loop radiative corrections to $\Gamma^{b}_{L,R}$ are given by the five triangle diagrams and two bottom-wave function renormalization diagrams. 
The triangle diagrams contributions to $\Gamma_{L,R}$ are given by
\beq
\parbox[c]{15ex}{\includegraphics[width=15ex]{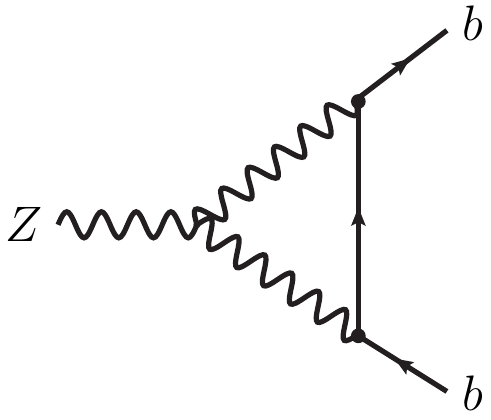}}
\equiv \hspace*{1ex}
\delta \Gamma^{(a)}_L
= \hspace*{1ex}
\frac{-g^2}{32 \pi^2} \cdot g^{(a)}_{ij}
\left[ 
-2q^2 \left( C_1 + C_2 +  C_{12} \right) + 12 C_{00}
\right] (0,0,q^2, m^2_t,M^2_{G^\pm_i},M^2_{G^\pm_j})\,,
\label{Zbb-1loop-Vtriangle-top-exchange}
\eeq
\beq
\parbox[c]{15ex}{\includegraphics[width=15ex]{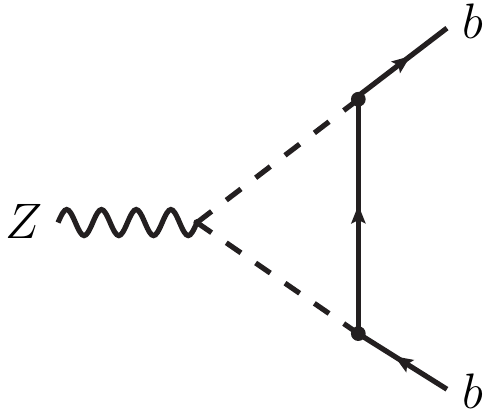}}
\equiv \hspace*{1ex}
\delta \Gamma^{(b)}_L
= \hspace*{1ex}
\frac{-1}{8 \pi^2}\cdot \frac{m^2_t}{f^2_\pi}\cdot g^{(b)}_{ij} C_{00}(0,0,q^2,m^2_t,M^2_{G^\pm_i},M^2_{G^\pm_j})\,,
\label{Zbb-1loop-NGBtriangle-top-exchange}
\eeq
\beq
\parbox[c]{15ex}{\includegraphics[width=15ex]{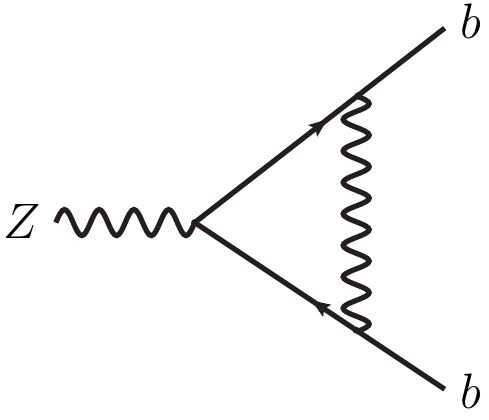}}
\equiv \hspace*{1ex}
\delta \Gamma^{(c)}_L
= \hspace*{1ex}
\frac{g^2}{32 \pi^2} 
\left[ 
\begin{aligned}
&g^{(c)}_{i L} \left\{2q^2 \left( C_1 + C_2 +  C_{11} + C_{12} \right) + 4 C_{00} \right\}
\\
&\hspace*{20ex} - 2 g^{(c)}_{i R} m^2_t C_0
\end{aligned}
\right] (q^2, 0,0,m^2_t,m^2_t,M^2_{G^\pm_i})
\,,
\label{Zbb-1loop-Vtriangle-V-exchange}
\eeq
\beq
\parbox[c]{15ex}{\includegraphics[width=15ex]{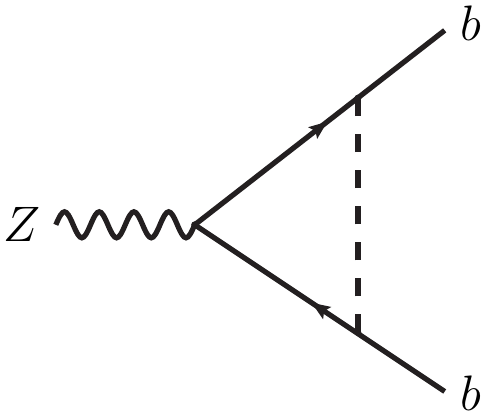}}
\equiv \hspace*{1ex}
\delta \Gamma^{(d)}_L
= \hspace*{1ex}
\frac{-1}{16 \pi^2} \cdot  \frac{m^2_t}{f^2_\pi}
\left[ 
\begin{aligned}
&-g^{(d)}_{i R} \left\{q^2 \left( C_1 +  C_{11} + C_{12} \right) + 2 C_{00} \right\}
\\
&\hspace*{20ex} +g^{(d)}_{i L} m^2_t C_0
\end{aligned}
\right] (q^2, 0,0,m^2_t,m^2_t,M^2_{G^\pm_i})
\,,
\label{Zbb-1loop-NGBtriangle-NGB-exchange}
\eeq
\beq
\parbox[l]{15ex}{\includegraphics[width=15ex]{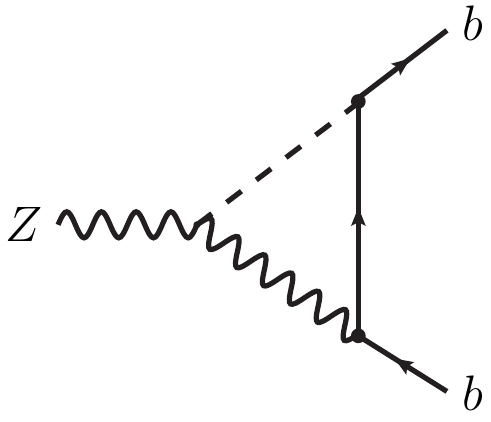}}
\hspace*{2ex} +\hspace*{2ex}
\parbox[r]{15ex}{\includegraphics[width=15ex]{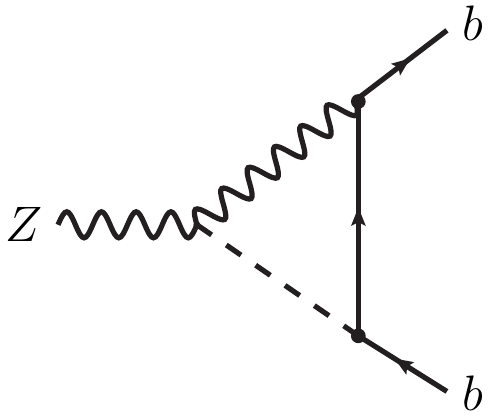}}
\equiv \hspace*{1ex}
\delta \Gamma^{(e)}_L
= \hspace*{1ex}
\frac{-g}{8 \pi^2}  \frac{m^2_t}{\sqrt{2}f_\pi} g^{(e)}_{ij} \cdot C_0 (0,q^2,0,m^2_t,M^2_{G^\pm_i},M^2_{G^\pm_j})
\,.
\label{Zbb-1loop-VNGBtriangle}
\eeq
The bottom-wave function renormalization diagram contributions to $\Gamma_{L,R}$ are given by 
\beq
\parbox[l]{15ex}{\includegraphics[width=15ex]{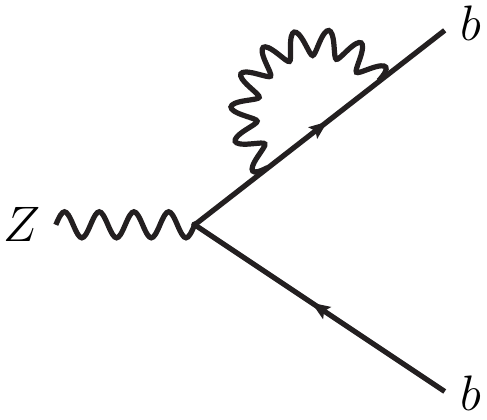}}
\hspace*{2ex} +\hspace*{2ex}
\parbox[r]{15ex}{\includegraphics[width=15ex]{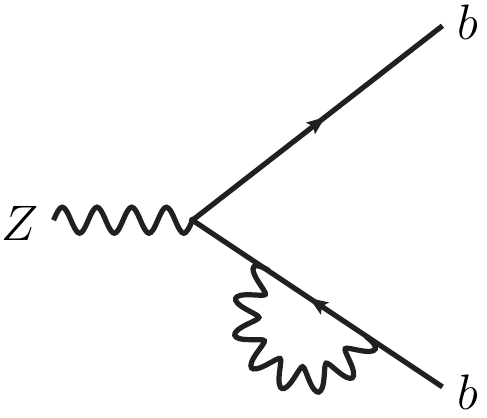}}
\equiv \hspace*{1ex}
\delta \Gamma^{(f)}_L
= \hspace*{1ex}
\frac{-g^2}{16 \pi^2} \cdot  g^{(f)}_i \cdot [B_0 (0, m^2_t,M^2_{G^\pm_i}) + B_1(0, m^2_t,M^2_{G^\pm_i})] 
\,,\label{Zbb-1loop-Wfn-V}
\eeq
\beq
\parbox[l]{15ex}{\includegraphics[width=15ex]{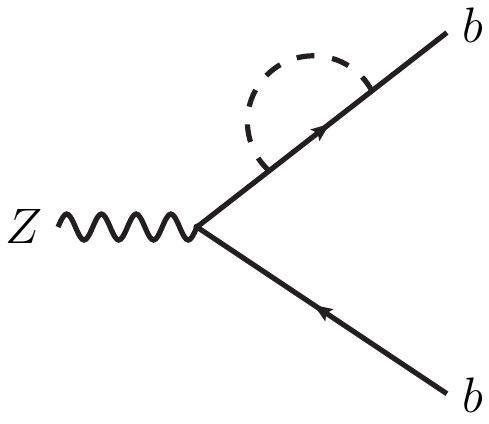}}
\hspace*{2ex} +\hspace*{2ex}
\parbox[r]{15ex}{\includegraphics[width=15ex]{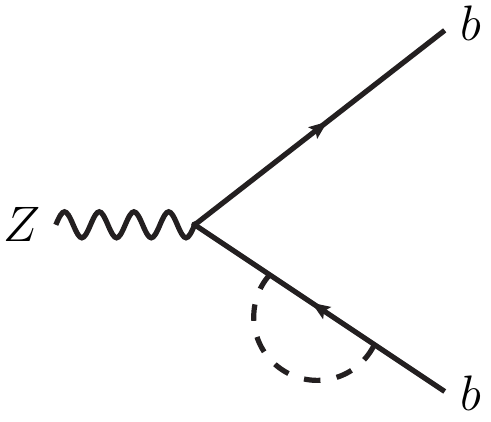}}
\equiv \hspace*{1ex}
\delta \Gamma^{(g)}_L
= \hspace*{1ex}
\frac{-1}{16 \pi^2} \cdot \frac{m^2_t}{f^2_\pi}  \cdot g^{(g)}_i \cdot [B_0 (0, m^2_t,M^2_{G^\pm_i}) + B_1(0, m^2_t,M^2_{G^\pm_i})] 
\,.\label{Zbb-1loop-Wfn-NGB}
\eeq
Here $q$ is short for a momentum of incoming $Z$ boson, and in the above results summation over the repetitive indices $i,j$ is implied. In Eqs.(\ref{Zbb-1loop-Wfn-V}) and (\ref{Zbb-1loop-Wfn-NGB}) we have taken into consideration the factor $1/2$ arising from wave function renormalization constant $Z^{1/2}_3$ of the bottom quark. The expressions for $g^{(k)}\,,(k=a,\cdots, g)$ in the present case are given in Table.\ref{Zbbvertex-coupling}.
In Eqs.(\ref{Zbb-1loop-Vtriangle-top-exchange})
-(\ref{Zbb-1loop-Wfn-NGB}), the $B$- and $C$-functions are,respectively, the two- and three-point loop integrals \cite{Denner:1991kt}. %
\begin{table}
{
\tabcolsep=1ex
\renewcommand\arraystretch{2}
\begin{tabular}{|c|c|}
\hline 
$g^{(a)}_{ij}$& $\kappa^{(Z)}_{ij}\left[v^1_{G^\pm_i}\right]\Big[v^1_{G^\pm_j}\Big]$
\\ \hline
$g^{(b)}_{ij}$& $\lambda^{(Z)}_{ij}
 \left[v^1_{\pi^\pm_i}\right]\Big[v^1_{\pi^\pm_j}\Big]$
\\ \hline
$g^{(c)}_{i L(R)}$& $\{ (2/3)gs_\theta [v^0_Z] + (g/c_\theta) g^t_{L(R)} [v^1_Z] \}\Big[v^1_{G^\pm_i} \Big]^2$
\\ \hline
$g^{(d)}_{i L(R)}$& $\{ (2/3)gs_\theta [v^0_Z] + (g/c_\theta) g^t_{L(R)} [v^1_Z] \}\Big[v^1_{\pi^\pm_i}\Big]^2$
\\ \hline
$g^{(e)}_{ij}$& $ -g^{(Z)}_{ij}
\left[v^1_{\pi^\pm_i}\right]\Big[v^1_{G^\pm_j}\Big]$
\\ \hline
$g^{(f)}_i$& $\left\{ (-1/3)gs_\theta [v^0_Z] + (g/c_\theta) g^b_{L} [v^1_Z] \right\} \Big[v^1_{G^\pm_i}\Big]^2$
\\ \hline
$g^{(g)}_i$& $\left\{ (-1/3)gs_\theta [v^0_Z] + (g/c_\theta) g^b_{L} [v^1_Z] \right\}  \Big[v^1_{\pi^\pm_i}\Big]^2$
\\ \hline
\end{tabular}
}
\caption{
Couplings in Eqs.(\ref{Zbb-1loop-Vtriangle-top-exchange})
-(\ref{Zbb-1loop-Wfn-NGB}). The indices take values $i,j=1,2,3$ and $n=0,1,2,3$.
\label{Zbbvertex-coupling}
}
\end{table}
The concrete expressions are obtained from
\beq
B_{0;\mu} (p^2_1,m^2_1,m^2_2)
\!\!&=&\!\!
\frac{(2\pi\mu)^{4-d}}{i\pi^2}
\int \!\!\!d^d k \,\frac{1 ; k_\mu}{[k^2 - m^2_1][(k + p_1)^2 - m^2_2]}
\,,
\\
C_{0;\mu;\mu\nu} (p^2_1,p^2_2,(p_1+p_2)^2,m^2_1,m^2_2,m^2_3)
\!\!&=&\!\!
\frac{(2\pi\mu)^{4-d}}{i\pi^2}
\int \!\!\!d^d k \,\frac{1 ; k_\mu ; k_\mu k_\nu}{[k^2 - m^2_1][(k + p_1)^2 - m^2_2][(k + p_1 + p_2)^2 - m^2_3]}
\,,
\eeq
where $\mu$ is the regularization mass, via the Lorentz decomposition forms which are
\beq
B_\mu 
\!\!&=&\!\! 
p_{1\mu} \cdot B_1 
\,,\\
C_\mu 
\!\!&=&\!\! 
p_{1\mu} \cdot C_1 +  (p_1 + p_2)_\mu \cdot C_2
\,,\\
C_{\mu \nu} 
\!\!&=&\!\! 
p_{1\mu} p_{1\nu} \cdot C_{11} +  (p_1 + p_2)_\mu (p_1 + p_2)_\nu \cdot C_{22}
+ \{ p_1, p_1 + p_2\}_{\mu \nu} \cdot C_{12} + g_{\mu\nu} \cdot C_{00}
\,,\\
\{ P, Q\}_{\mu \nu}
\!\!&=&\!\!
P_\mu Q_\nu + Q_\mu P_\nu 
\,.
\eeq
The ultraviolet (UV) divergences appear only in $B_{0,1}, C_{00}$. In the dimensional regularization
these divergences are given by
\beq
4 C_{00}|_{\text{div.}}
=
B_0|_{\text{div.}}= -2 B_1|_{\text{div.}}
= \frac{1}{\bar{\epsilon}} \,,
\quad \text{where} \quad
\frac{1}{\bar{\epsilon}}=\frac{2}{4-d} - \gamma_E + \ln 4\pi
\,.
\eeq

In order to renormalize the UV  divergence in each diagram, we renormalize in accordance with 
\cite{Pich:2012sx}, i.e. we define the renormalized form factors as
\beq
F^{(k)}_{L}(m^2_t,m^2,M^2) \equiv 
\delta \Gamma^{(k)}_{L}(m^2_t,m^2,M^2) 
- \delta \Gamma^{(k)}_{L}(0,m^2,M^2) \,.
\label{Pich-renormalization}
\eeq
%
%
%
Thus, the UV-finite $\Gamma^b_{L,R}$ in the present GHLS case are 
\beq
\Gamma^b_L[\text{GHLS}] = \Gamma^b_L \left[ \text{GHLS,tree} \right] +\sum_k F^{(k)}_L(m^2_t,m^2,M^2)
\quad , \quad
\Gamma^b_R[\text{GHLS}] = \Gamma^b_R \left[ \text{GHLS,tree} \right] 
\,.
\eeq
To see the differences of $\Gamma^b_{L,R}$ between the present case and the SM case, we define 
$[\Delta \Gamma^b_{L,R}]_{\rm NP}$ as
\beq
\left[ \Delta \Gamma^b_{L,R}\right]_{\rm NP}
\equiv
\frac{g_{\rm EW}}{c_W } \left[ \delta g^b_{L,R} \right]_{\rm NP}
\equiv
 \Gamma^b_{L,R} \left[ \text{GHLS} \right]  - \Gamma^b_{L,R}[\text{SM}] 
\,,
\eeq
where $\Gamma^b_{L,R}[\text{SM}]$ is obtainable from the leading term of $ \Gamma^b_{L,R}[\text{GHLS}] $after expanding in $\epsilon_W \equiv g_{\rm EW}/\tilde{g}$ instead of $\epsilon= g/\tilde{g}$.

To obtain the dominant contribution from the one loop computation, we consider the ${\cal O}(\epsilon^0)$ terms in the expansion of $g^{(k)}_L$ to clearly see the results from the present model. First, we obtain the  contributions
\beq
&&
g^{(a)}_{11} = g_{\rm EW} c_W [1 + {\cal O}(\epsilon^2_W)]
\,, \quad
g^{(b)}_{11}=\frac{g_{\rm EW} (c^2_W - s^2_W)}{2c_W} [1 + {\cal O}(\epsilon^2_W)]
\,, \\[1ex]
&&
g^{(c)}_{1 L(R)}= g^{(d)}_{1 L(R)} = \frac{g_{\rm EW}}{c_W} g^t_{L(R)} [1 + {\cal O}(\epsilon^2_W)]
\,,
\\[1ex]
&&
g^{(e)}_{11} = \frac{g^2_{\rm EW}  s^2_W f_\pi}{\sqrt{2} c_W} [1 + {\cal O}(\epsilon^2_W)]
\,,\\[1ex]
&&
g^{(f)}_1= g^{(g)}_1 = \frac{g_{\rm EW}}{c_W} g^b_L [1 + {\cal O}(\epsilon^2_W)]
\,,
\eeq
which reproduce the SM one loop results at ${\cal{O}}(\epsilon_W^0)$.
Almost all other couplings are ${\cal O}(\epsilon^2_W)$. The only exception is $g^{(e)}_{12}$ which  starts with ${\cal O}(\epsilon^0_W)$ term given by
\beq
g^{(e)}_{12}
=
\frac{g^2_{\rm EW} s^2_W f_\pi}{\sqrt{2} c_W}
\left[ \frac{-(1-\chi^2)f^2_\sigma}{2 s^2_W f^2_\pi}  + {\cal O}(\epsilon^2_W) \right]
\,.
\eeq
Hence this $g^{(e)}_{12}$ is the leading one loop contribution beyond the SM contribution to $R_b$ in the GHLS case. Note that this 
${\cal O}(\epsilon^0_W)$ contribution vanishes if $\chi^2 \to 1$. Thus the leading radiative contribution to 
$[\delta g^b_L]_{\rm NP}$ is given by
\beq
\left[ \delta g^b_L \right]^{\rm 1 loop}_{\rm NP}
=
 \frac{g^2_{\rm EW}  m^2_t}{16 \pi^2} 
 \frac{1-\chi^2}{2}
 \frac{f^2_\sigma}{f^2_\pi} \cdot C_0(0,q^2,0,m^2_t,M^2_W,M^2_{V^\pm})
\left[ 1 + {\cal O}(\epsilon^2_W)\right]
\,,\label{1-loop-approx}
\eeq
where $C_0(0,q^2,0,m^2_1,m^2_2,M^2)$ with large $M^2$ is given by
\beq
C_0(0,q^2,0,m^2_1,m^2_2,M^2) 
= 
\frac{1}{M^2}
 \left[
  \frac{m^2_1}{m^2_1- m^2_2} \ln \frac{m^2_1}{M^2}
  -
 \frac{m^2_2}{m^2_1 - m^2_2} \ln \frac{m^2_2}{M^2}
 \right]
 \quad , \quad
 \text{for \,\,\,$\dfrac{q^2}{M^2},\dfrac{m^2_1}{M^2},\dfrac{m^2_2}{M^2} \ll 1$}
 \,.\label{C0-approx}
\eeq
Thus we obtain the result that the leading contribution, of the order ${\cal O}(\epsilon^0_W)$, due to vector 
mesons reproduces the SM contribution 
and at the same order gives a new physics contribution proportional to $1-\chi^2$.  

Let us then consider the $R_b$ constraint for the GHLS case. It is convenient to divide $R_b$ into two parts as
\beq
R_b = R^{\rm SM}_b + \Delta R_b \,, 
\eeq
where $\Delta R_b$ is defined as
\beq
\Delta R_b = 2 R^{\rm SM}_b (1-R^{\rm SM}_b) 
{\rm Re} \left[
\frac{g^b_L \left[ \delta g^b_L\right]_{\rm NP} + g^b_R \left[ \delta g^b_R\right]_{\rm NP}}{(g^b_L)^2 + (g^b_R)^2}
\right]
\label{dRb}
\,.
\eeq
With these definitions, naturally $\Delta R_b(\text{SM}) =0$. The experimental value of 
$R_b$ \cite{Baak:2011ze} is
\beq
R^{\rm exp}_b  \equiv \frac{\Gamma(Z \to \bar{b}b)}{\Gamma(Z \to {\rm had})}=0.21629 \pm 0.00066 \,.
\label{Rb-exp}
\eeq
and $R^{\rm SM}_b$ is the SM value which is predicted by the electroweak fit  as \cite{Baak:2011ze}
\beq
R^{\rm SM}_b = 0.21578 ^{+ 0.00005}_{-0.00008} \,.
\eeq
Thus the constraint for $\Delta R_b$ is given by
\beq
\Delta R_b = 0.00051 \pm 0.00066\,,
\label{Rb-NPconstraint}
\eeq
where the error is taken from the experimental error in Eq. (\ref{Rb-exp}). 

In addition, we impose the first Weinberg sum rule and the representation of $S$ in the GHLS case,
\beq
f^2_\sigma = f^2_\pi + \chi^2 f^2_q
\quad , \quad
S = \frac{8\pi}{\tilde{g}^2} (1-\chi^2)\,,
\eeq
which shows that the limit $\chi^2 \to 1$ corresponds with $S \to 0$, and in this limit the SM result is recovered.

\begin{figure}[htbp]
\begin{center}
\includegraphics[scale=0.9]{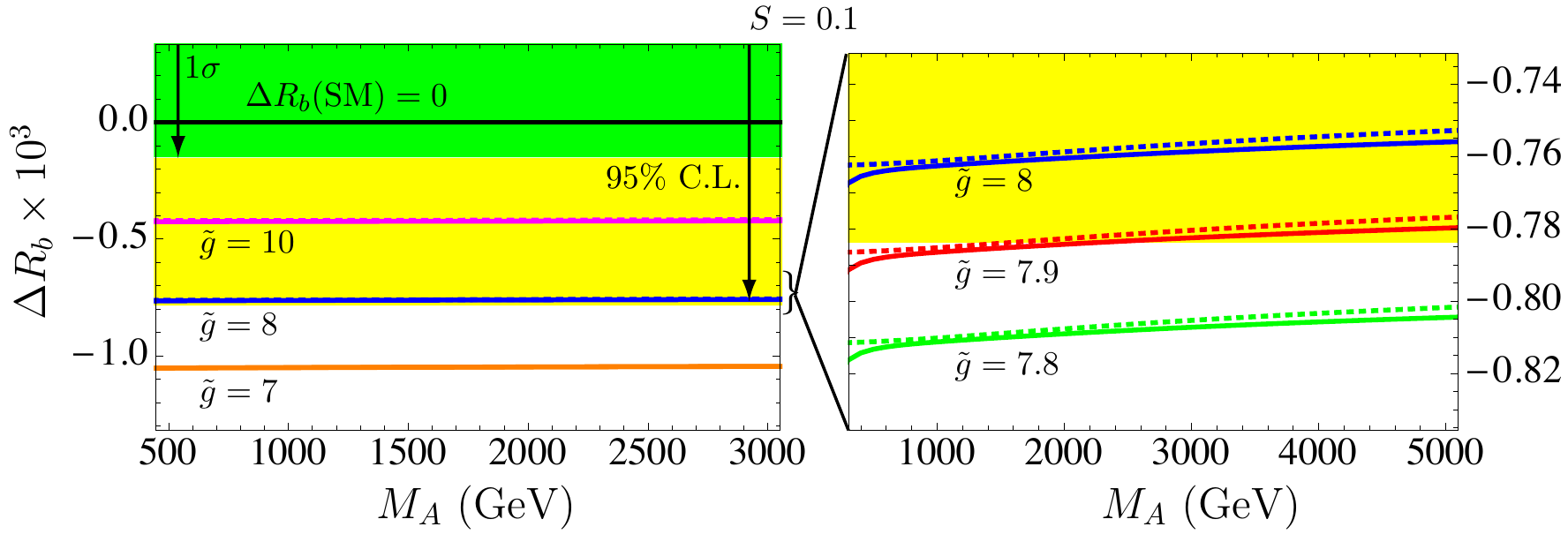} 
\\[2ex]
\includegraphics[scale=0.9]{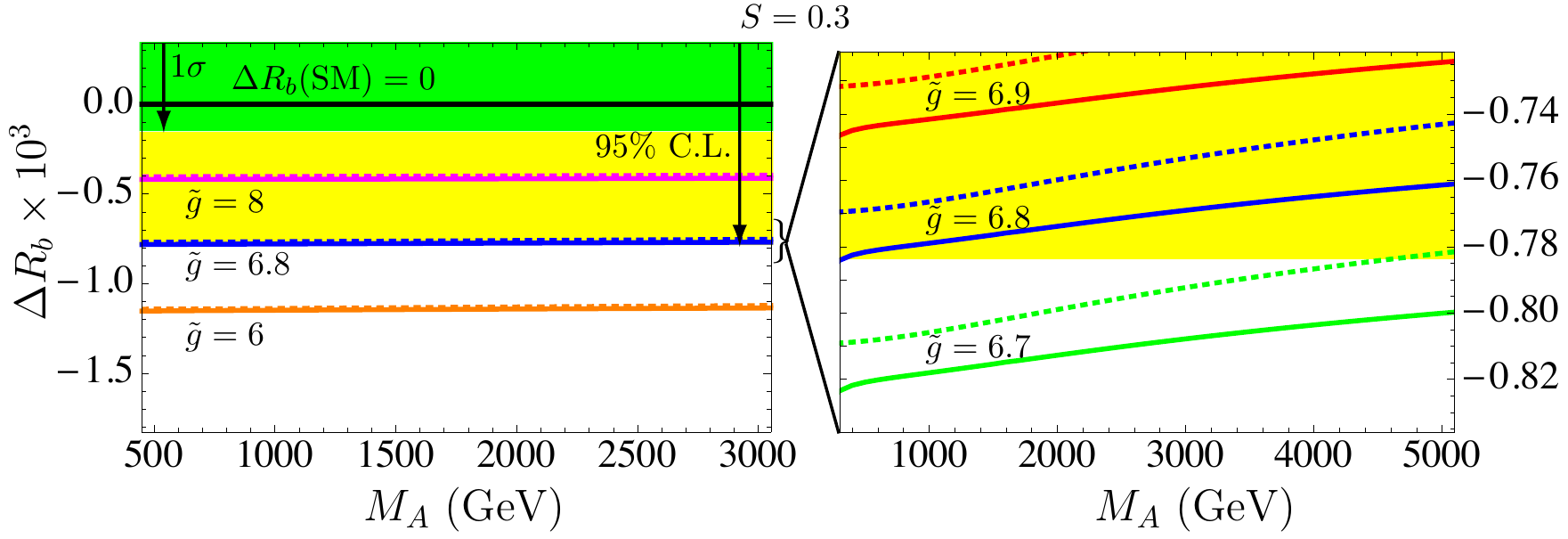} 
\caption{ 
$\Delta R_b$ as a function of $M_A$ for several values of $\tilde{g}$ for $S = 0.1$ (upper panels) and $S=0.3$ (lower panels) together with $1\sigma$,$95\%\cl$ constraints for $R_b$. The upmost solid lines show $\Delta R_b(\text{SM}) =0$. The dotted lines are results with Eq.(\ref{1-loop-approx}) and the solid lines are results with the {\it FormCalc/LoopTools} implementation of the present GHLS model. The right panels focus around the $95\%\cl$ lower constraint for $\Delta R_b$. 
\label{Rb-MA-constraints}}
\end{center}
\end{figure}

In Fig.\ref{Rb-MA-constraints}, we show the $\Delta R_b$ as a function of $M_A$ for $S=0.1,0.3$ and several values of $\tilde{g}$. The dotted lines in Fig.\ref{Rb-MA-constraints}, correspond to keeping only 
${\cal O}(\epsilon^0_W)$ terms in the one loop results, while the solid lines correspond to keeping all terms 
up to ${\cal O}(\epsilon^2_W)$ in the one loop results. We can see that considering only 
Eq.(\ref{1-loop-approx}) indeed gives the leading radiative correction. To obtain the results depicted by the solid lines, we evaluated $\delta g^b_L$ by using {\it FormCalc/LoopTools} with implementation of the present GHLS case with $q^2 = M^2_Z$. As a simple check on our numerical results, we can compare  
with the well known SM results. The SM limit from our one loop result is
\beq
\delta g^b_L [\text{GHLS}\rightarrow\text{ SM}] = 0.00308\,,
\eeq
for $s^2_W = 0.23$, $m_t = 172.9 \GeV$, $G_F = 1.16 \times 10^{-5} \GeV^{-2}$ and $M_Z = 91.2 \GeV$. This is consistent with the result $\delta g^b_L [\text{SM 1-loop}] = 0.0036$ in \cite{Bernabeu:1987me} and the result in the gaugeless limit e.g. in \cite{Abe:2009ni}, $\delta g^L_b = m^2_t/(16 \pi^2 v^2_{\rm EW}) \simeq 0.0032$. From Fig.\ref{Rb-MA-constraints}, we conclude that for wide range of masses $M_A $, 
the coupling $\tilde{g}$ must be large. The precise bounding value depends on the value of the $S$ parameter, e.g.
\beq
\begin{aligned}
\tilde{g} &\gtrsim& 8 &\quad \text{for $S=0.1$}\,, \\[1ex]
\tilde{g} &\gtrsim& 7 &\quad \text{for $S=0.3$}\,.
\end{aligned} 
\eeq

\begin{figure}[htbp]
\begin{center}
\includegraphics[scale=0.9]{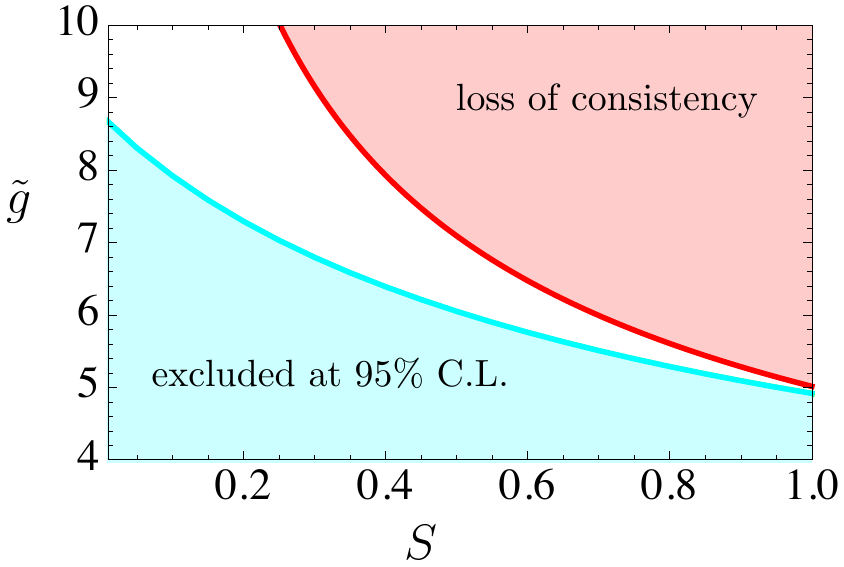}
\caption{ 
Constraints on $(S,\tilde{g})$-plane. The curves are practically independent of $M_A$. In the shaded region in the top-right corner (red) the model is inconsistet as $\chi ^2 <0$ there. The lower shaded region is 
excluded by the $95 \% \cl$ on $\Delta R_b$.
\label{Rb-gtS}}
\end{center}
\end{figure}

In Fig.\ref{Rb-gtS}, we show the constraint $\Delta R_b$ in Fig.\ref{Rb-MA-constraints} on $(S,\tilde{g})$ plane for several values of $M_A$. The shaded region in the top-right corner is excluded by consistency: $\chi^2$ is positive only below this region. The shaded region in the lower-right corner is excluded by the constraints on $\Delta R_b$ as shown in Fig. \ref{Rb-MA-constraints}. From Fig.\ref{Rb-gtS}, we find that the existence of light axial vector meson with small $\tilde{g}$, i.e. $\tilde{g} \lessim 6$ is not allowed and this results should be taken into consideration for the phenomenology based on the GHLS. 

\section{Conclusion}
\label{checkout}

We have studied a generic effective Lagrangian for dynamical electroweak symmetry breaking assuming global symmetry breaking pattern SU(2)$\times$SU(2). Including the composite vector states and assuming 
minimal coupling to the Standard Model fields, we have evaluated the $R_b$ constraints.  After imposing the Weinberg sum rules on the model parameters, we studied the $R_b$ constraints in the ($M_A,\tilde{g}$)-plane. The constraints were shown to lead to only very mild dependence of $\tilde{g}$ on $M_A$, and the robust result is that for any $M_A$, a sufficiently strong coupling $\tilde{g}\sim{\cal{O}}(10)$ allows one to saturate the $R_b$ constraints. The results are also dependent on the value of the precision parameter $S$, and we also observed that small values of $S$ also imply large $\tilde{g}$.

The models featuring new strong dynamics responsible for the electroweak symmetry breaking are compatible with the current experimental data and provide viable candidates for physics beyond the Standard Model. Our study was carried out for the GHLS type non-linear sigma model Lagrangian with a minimal coupling to SM flavors, and our result should be applicable and useful for any model where such strongly intracting sector appears. To make our results easily applicable,  we identified the coupling giving the dominant contribution to $\Delta R_b$. In the case of nonminimal coupling between the strongly interacting fields and SM matter requires a further analysis, which we leave for future work.

\appendix
\def\thesection{\Alph{section}}
\renewcommand{\theequation}{\Alph{section}.\arabic{equation}}
\setcounter{equation}{0}

\section{Diagonalization}
\label{app-diag}

The mass terms are given in Eq.(\ref{mass-GHLS}), and we expand each representation in 
$\epsilon=g/\tilde{g}$. 

\subsection{Charged gauge sector}
\setcounter{equation}{0}
\numberwithin{equation}{section}

In Eq.(\ref{mass-GHLS}), the mass term for the charged sector is given by 
\beq
{\cal L}_{CS} =
(\tilde{W}^-_\mu \, \tilde{V}^-_\mu \, \tilde{A}^-_\mu) \cdot \tilde{{\cal M}}^2_{{\cal G}CC} \cdot (\tilde{W}^{+\mu} \, \tilde{V}^{+\mu} \, \tilde{A}^{+\mu})^T\,,
\eeq
where 
\beq
\tilde{{\cal M}}^2_{{\cal G}CC} 
\!\!\!&=&\!\!\!
\bpm
\dfrac{1}{2}g^2 \!\left[ f^2_\pi + f^2_\sigma + \chi^2 f^2_q \right]  
& -\dfrac{1}{\sqrt{2}} g\tilde{g} f^2_\sigma & -\dfrac{\chi}{\sqrt{2}} g\tilde{g} f^2_q \\[3ex]
-\dfrac{	1}{\sqrt{2}} g\tilde{g} f^2_\sigma & \tilde{g}^2 f^2_\sigma & 0 \\[3ex]
-\dfrac{\chi}{\sqrt{2}} g\tilde{g} f^2_q & 0 &  \tilde{g}^2 f^2_q
\epm\,. \label{GCC-mass}
\eeq
The diagonalizing matrix ${\cal O}_{{\cal G}CC}$ for $\tilde{{\cal M}}^2_{{\cal G}CC}$ is
\beq
O_{{\cal G}CC} \equiv
\bpm
v^1_W & v^1_{V^\pm} & v^1_{A^\pm}\\[1ex]
v^2_W & v^2_{V^\pm} & v^2_{A^\pm} \\[1ex]
v^3_W & v^3_{V^\pm} & v^3_{A^\pm}
\epm\,,
\eeq
which is an orthonormal matrix,  and then ${\cal M}^2_{{\cal G}CC} ={\cal O}^T_{{\cal G}CC} \tilde{\cal M}^2_{{\cal G}CC} {\cal O}^{}_{{\cal G}CC}$ is a diagonal matrix $\text{diag}\,(M^2_W,M^2_{V^\pm},M^2_{A^\pm})$ where each eigenvalue is given in Eq.(\ref{eigenvalues-GCC}). 
the quantities $v^i_W$ in ${\cal O}_{{\cal G}CC}$ are given by
\beq
v^1_W = 1 - \frac{1+\chi^2}{4} \epsilon^2 + {\cal O}(\epsilon^4)
\,,\quad
v^2_W = \frac{1}{\sqrt{2}} \epsilon + {\cal O}(\epsilon^4)
\,,\quad
v^3_W = \frac{\chi}{\sqrt{2}} \epsilon + {\cal O}(\epsilon^4)
\,,
\eeq
$v^i_{V^\pm}$ in ${\cal O}_{{\cal G}CC}$ are given by
\beq
v^1_{V^\pm} = -\frac{1}{\sqrt{2}} \epsilon + {\cal O}(\epsilon^4) 
\,,\quad
v^2_{V^\pm} = 1 -\frac{\epsilon ^2}{4} + {\cal O}(\epsilon^4)
\,,\quad
v^3_{V^\pm} = \frac{\chi  f_q^2}{2 (f_{\sigma }^2 - f_q^2)} \epsilon^2+ {\cal O}(\epsilon^4)
\,,
\eeq
and finally $v^i_{A^\pm}$ in ${\cal O}_{{\cal G}CC}$ are given by
\beq
v^1_{A^\pm} =  -\frac{1 }{\sqrt{2}} \epsilon + {\cal O}(\epsilon^4)
\,,\quad
v^2_{A^\pm} = \frac{\chi  f_{\sigma }^2}{2 (f_q^2-f_{\sigma }^2)}\epsilon^2 + {\cal O}(\epsilon^4)
\,,\quad
v^3_{A^\pm} =1 -\frac{\chi ^2}{4} \epsilon ^2 + {\cal O}(\epsilon^4)
\,.
\eeq
%

\subsection{Neutral gauge sector}
\beq
\tilde{{\cal M}}^2_{{\cal G}NC} 
\!\!\!&=&\!\!\!
\bpm
\dfrac{1}{2}g'^2 \!\left[ f^2_\pi + f^2_\sigma + \chi^2 f^2_q \right] 
& \dfrac{1}{2}g'g \!\left[ -f^2_\pi + f^2_\sigma - \chi^2 f^2_q \right]
& -\dfrac{1}{\sqrt{2}} g'\tilde{g} f^2_\sigma & \dfrac{\chi}{\sqrt{2}} g'\tilde{g} f^2_q 
\\[3ex]
\dfrac{1}{2}g'g \!\left[ -f^2_\pi + f^2_\sigma - \chi^2 f^2_q \right]
& \dfrac{1}{2}g^2 \!\left[ f^2_\pi + f^2_\sigma + \chi^2 f^2_q \right]  
& -\dfrac{1}{\sqrt{2}} g\tilde{g} f^2_\sigma & -\dfrac{\chi}{\sqrt{2}} g\tilde{g} f^2_q 
\\[3ex]
-\dfrac{1}{\sqrt{2}} g'\tilde{g} f^2_\sigma & -\dfrac{1}{\sqrt{2}} g\tilde{g} f^2_\sigma 
& \tilde{g}^2 f^2_\sigma & 0 
\\[3ex]
\dfrac{\chi}{\sqrt{2}} g'\tilde{g} f^2_q & -\dfrac{\chi}{\sqrt{2}} g\tilde{g} f^2_q & 0 
&  \tilde{g}^2 f^2_q
\epm\,,\label{NC-mass}
\eeq

The diagonalizing  matrix $O_{{\cal G}NC}$ for $\tilde{{\cal M}}^2_{{\cal G}NC} $ is
\beq
O_{{\cal G}NC}\equiv
\bpm
v^0_\gamma & v^0_Z & v^0_{V^0} & v^0_{A^0} \\[1ex]
v^1_\gamma & v^1_Z & v^1_{V^0} & v^1_{A^0} \\[1ex]
v^2_\gamma & v^2_Z & v^2_{V^0} & v^2_{A^0} \\[1ex]
v^3_\gamma & v^3_Z & v^3_{V^0} & v^3_{A^0}
\epm\,,
\eeq
 and then ${\cal M}^2_{{\cal G}NC} ={\cal O}^T_{{\cal G}NC} \tilde{\cal M}^2_{{\cal G}NC} {\cal O}^{}_{{\cal G}NC}$ is a diagonal matrix $\text{diag\,}(0,M^2_Z,M^2_V,M^2_A)$ where each eigenvalue is given in Eq.(\ref{eigenvalues-GNC}). 
%
The quantities $v^i_\gamma$ of ${\cal O}_{{\cal G}NC}$ are given by
\beq
v^0_\gamma = 1 -  s^2_\theta \epsilon^2 +{\cal O}(\epsilon^4)
\,, \quad
v^1_\gamma =0
\,, \quad
v^2_\gamma
=\sqrt{2} s_\theta v^0_\gamma \epsilon 
\,, \quad
v^3_\gamma = 0
\,,
\eeq
$v^i_Z$ of ${\cal O}_{{\cal G}NC}$ are given by
\beq
&&
v^0_Z =
-  c_{2 \theta}\, t_\theta \epsilon^2 +{\cal O}(\epsilon^4)
\,,\quad
v^1_Z =
1 -\frac{c^2_{2\theta}+\chi^2}{4 c^2_\theta} \epsilon^2 +{\cal O}(\epsilon^4)
\,, \nonumber\\[1ex]
&&
v^2_Z = \frac{c_{2\theta}}{\sqrt{2}c_\theta} \epsilon +{\cal O}(\epsilon^3)
\,,\quad
v^3_Z = \frac{\chi}{\sqrt{2}c_\theta} \epsilon +{\cal O}(\epsilon^3)
\,,\label{ev-Z}
\eeq
$v^i_{V^0}$ of ${\cal O}_{{\cal G}NC}$ are given by
\beq
&&
v^0_{V^0} = -\sqrt{2} s_\theta \epsilon +{\cal O}(\epsilon^4)
\,, \quad
v^1_{V^0} = -\frac{c_{2\theta}}{\sqrt{2}c_\theta} \epsilon +{\cal O}(\epsilon^4) 
\,,\nonumber\\[1ex]
&&
v^2_{V^0} = 1 - \dfrac{1}{4c^2_\theta} \epsilon^2 +{\cal O}(\epsilon^4)
\,,\quad
v^3_{V^0} =
\dfrac{c_{2\theta} \chi f^2_q}{2c^2_\theta (f^2_\sigma - f^2_q)}  \epsilon^2 +{\cal O}(\epsilon^4)
\,,
\eeq
$v^i_{A^0}$ of ${\cal O}_{{\cal G}NC}$ are given by
\beq
&&
v^0_{A^0} = 0 +{\cal O}(\epsilon^4)
\,,\quad
v^1_{A^0} = -\frac{\chi}{\sqrt{2}c_\theta} \epsilon +{\cal O}(\epsilon^4)
\,,\nonumber\\[1ex]
&&
v^2_{A^0} = 
\frac{c_{2\theta}\chi f^2_\sigma}{2c^2_\theta\left( f^2_q - f^2_\sigma \right)} \epsilon^2 +{\cal O}(\epsilon^4)
\,,\quad 
v^3_{A^0} = 1 - \frac{\chi^2}{4c^2_\theta} \epsilon^2 +{\cal O}(\epsilon^4)\,,
\eeq
Here $\tan \theta = g'/g$ and we denoted $s_\theta \equiv \sin \theta, s_{2\theta} \equiv \sin 2 \theta\,$ etc. for simplicity.

\subsection{Charged NGB sector}
\numberwithin{equation}{section}
We should note that all NGBs are absorbed by the gauge bosons, i.e. eigenvalues of $\tilde{\cal M}^2_{\Sigma}$ should be equal to the eigenvalues of $\tilde{\cal M}^2_{\cal G}$ in the Feynman gauge. The diagonalizing matrix ${\cal O}_{{\cal G}CC}$ for $\tilde{{\cal M}}^2_{{\cal G}CC}$ is
\beq
{\cal O}_{\Sigma CC} \equiv
\bpm
v^1_{\pi^\pm} & v^1_{\pi^\pm_\sigma} & v^1_{\pi^\pm_q}\\[1ex]
v^2_{\pi^\pm} & v^2_{\pi^\pm_\sigma} & v^2_{\pi^\pm_q} \\[1ex]
v^3_{\pi^\pm} & v^3_{\pi^\pm_\sigma}& v^3_{\pi^\pm_q}
\epm\,,
\eeq
where $v^i_{\pi^\pm}$ of ${\cal O}_{\Sigma CC}$ are given by
\beq
v^1_{\pi^\pm} = 1 + {\cal O}(\epsilon^4)
\,,\quad
v^2_{\pi^\pm} = \frac{f_\pi}{2f_\sigma} \epsilon ^2 + {\cal O}(\epsilon^4)
\,,\quad
v^3_{\pi^\pm} = -\frac{\chi f_\pi}{2f_q} \epsilon ^2 + {\cal O}(\epsilon^4)\,,
\eeq
$v^i_{\pi^\pm_\sigma}$ of ${\cal O}_{\Sigma CC}$ are given by
\beq
v^1_{\pi_\sigma^\pm} = -\frac{f_\pi}{2f_\sigma} \epsilon^2 + {\cal O}(\epsilon^4)
\,,\quad
v^2_{\pi_\sigma^\pm} = 1 + {\cal O}(\epsilon^4)
\,, \quad
v^3_{\pi_\sigma^\pm} = \frac{ \chi  f_q f_{\sigma }}{2 (f_q^2 - f_{\sigma }^2)}\epsilon ^2 + {\cal O} (\epsilon^4)
\,,
\eeq
$v^i_{\pi^\pm_q}$ of ${\cal O}_{\Sigma CC}$ are given by
\beq
v^1_{\pi_q^\pm} = - \frac{ \chi f_{\pi }}{2 f_q } \epsilon^2 + {\cal O} (\epsilon^4)
\,,\quad
v^2_{\pi_q^\pm} =\frac{\chi  f_q f_{\sigma }}{2 (f_q^2 - f_{\sigma }^2)} \epsilon ^2 + {\cal O} (\epsilon^4)
\,,\quad
v^3_{\pi_q^\pm} = -1 + {\cal O} (\epsilon^4)
\,,
\eeq

\subsection{Neutral NGB sector}
\numberwithin{equation}{section}

The diagonalizing matrix ${\cal O}_{{\cal G}NC}$ for $\tilde{{\cal M}}^2_{{\cal G}NC}$ is
\beq
{\cal O}_{\Sigma NC} \equiv
\bpm
v^1_{\pi^0} & v^1_{\pi^0_\sigma} & v^1_{\pi^0_q}\\[1ex]
v^2_{\pi^0} & v^2_{\pi^0_\sigma} & v^2_{\pi^0_q} \\[1ex]
v^3_{\pi^0} & v^3_{\pi^0_\sigma}& v^3_{\pi^0_q}
\epm\,,
\eeq
where $v^i_{\pi^0}$ of ${\cal O}_{\Sigma NC}$ are given by 
\beq
v^1_{\pi^0} = 1 + {\cal O} (\epsilon^4)
\,,\quad
v^2_{\pi^0} = \frac{ c_{2 \theta} f_{\pi } }{2 c^2_\theta f_{\sigma }} \epsilon ^2 + {\cal O} (\epsilon^4)
\,,\quad
v^3_{\pi^0} = -\frac{\chi f_{\pi }}{2 c^2_\theta f_q} \epsilon ^2 + {\cal O} (\epsilon^4)
\,,
\eeq
$v^i_{\pi^0_\sigma}$ of ${\cal O}_{\Sigma NC}$ are given by
\beq
v^1_{\pi^0_\sigma} = -\frac{c_{2\theta}}{c^2_\theta}
\frac{f_{\pi }}{2 f_{\sigma }} \epsilon ^2 + {\cal O} (\epsilon^4)
\,,\quad
v^2_{\pi^0_\sigma} = 1 + {\cal O} (\epsilon^4)
\,,\quad
v^3_{\pi^0_\sigma} =
\frac{c_{2\theta} \chi  f_q f_{\sigma } }{2 c^2_\theta (f_q^2 - f_{\sigma }^2)}  
\epsilon ^2 + {\cal O} (\epsilon^4)
\,,
\eeq
$v^i_{\pi^0_q}$ of ${\cal O}_{\Sigma NC}$ are given by
\beq
v^1_{\pi^0_q} = -\frac{\chi f_{\pi }}{2c^2_\theta f_q} \epsilon ^2 + {\cal O} (\epsilon^4)
\,,\quad
v^2_{\pi^0_q} = \frac{c_{2 \theta} \chi  f_q f_{\sigma }}{2 c^2_\theta \left(f_q^2-f_{\sigma }^2\right)}  \epsilon ^2 + {\cal O} (\epsilon^4)
\,,\quad
v^3_{\pi^0_q} = -1 + {\cal O} (\epsilon^4)
\,.
\eeq


\end{document}